\documentclass[journal]{IEEEtran}
\usepackage{amsmath,amsfonts}
\usepackage{array}
\usepackage[caption=false,font=normalsize,labelfont=sf,textfont=sf]{subfig}
\usepackage{textcomp}
\usepackage{stfloats}
\usepackage{url}
\usepackage{verbatim}
\usepackage{graphicx}
\usepackage{cite}

\usepackage[hidelinks]{hyperref}
\usepackage{newtxmath}

\graphicspath{{figures/}}

\begin{document}

\title{Constraint Hypergraphs as a Unifying Framework for Digital Twins}

\author{John Morris, Douglas L. Van Bossuyt, Edward Louis, Gregory Mocko, John Wagner%
\thanks{J. Morris, E. Louis, G. Mocko, and J. Wagner are with the Dept. of Mechanical Engineering at Clemson University, Clemson, SC, USA}%
\thanks{D. L. Van Bossuyt is with the Dept. of Systems Engineering at the Naval Postgraduate School, Monterey, CA, USA}%
\thanks{The views presented are those of the authors and do not necessarily represent the views of the U.S. Navy or any other organization. This material is declared a work of the U.S. Government and is not subject to copyright protection in the United States. Approved for Public Release; distribution is unlimited.}%
\thanks{Manuscript submitted 26 Sep 2025.}}

\markboth{Preprint}%
{Morris, Van Bossuyt, \MakeLowercase{\textit{et al.}}: Constraint Hypergraphs as a Unifying Framework for Digital Twins}

\IEEEpubid{This work has been submitted to the IEEE for possible publication. Copyright may be transferred without notice, after which this version may no longer be accessible.}

\maketitle

\begin{abstract}
    Digital twins, used to represent physical systems, have been lauded as tools for understanding reality. Complex system behavior is typically captured in domain-specific models crafted by subject experts. Contemporary methods for employing models in a digital twin require prescriptive interfaces, resulting in twins that are difficult to connect, redeploy, and modify. The limited interoperability of these twins has prompted calls for a universal framework enabling observability across model aggregations. Here we show how a new mathematical formalism called a constraint hypergraph addresses these challenges in interoperability. A digital twin is shown to be the second of two coupled systems where both adhere to the same constraint hypergraph, permitting the properties of the first to be observable from the second. Interoperability is given by deconstructing models into a structure enabling autonomous, white-box simulation of system properties. The resulting digital twins can interact immediately with both human and autonomous agents. This is demonstrated in a case study of a microgrid, showing how both measured and simulated data from the aggregated twins can be provided regardless of the operating environment. By connecting models, constraint hypergraphs supply scientists and modelers robust means to capture, communicate, and combine digital twins across all fields of study. We expect this framework to expand the use of digital twins, enriching scientific insights and collaborations by providing a structure for characterizing complex systems.
\end{abstract}


\section{Introduction}
\IEEEPARstart{E}{very} thing in the world is a system, and every scientific endeavor is fundamentally concerned with describing those systems and their behaviors. A system is a collection of things that affect the world in unique ways when arranged together \cite{internationalorganizationforstandardizationSystemsSoftwareEngineering2023}. Explanations of system behavior are expressed in modeling languages such as the algebraic models used to represent systems of numbers, hierarchical charts for organizations, and diagrams of electric circuits. By default, a model represents only a virtual system, meaning the system it describes exists solely as information. For example, a model of child development describes not a specific child, but a virtual child representing all children. But additional, specialized information can be revealed by building models that describe a single, physical system. Such specific representations are termed digital twins (DTs), and are employed to expose the complex interactions of real systems, allowing previously indiscernible relationships to be described. DTs leveraging modern data processing capabilities have been proposed for solving problems in both scientific and industrial communities such as medical diagnostics \cite{sahalPersonalDigitalTwin2022}, global weather forecasting \cite{hoffmannDestinationEarthDigital2023}, and machine maintenance \cite{fuDigitalTwinIntegration2022}. 

Because of the complexity of the systems they represent, high fidelity DTs end up being excruciatingly intricate. Their convolutions result in fragile representations that are difficult to maintain \cite{koskinenSoftwareMaintenanceCost2003}, resistant to reuse and redeployment \cite{zambranoIndustrialDigitalizationIndustry2022}, and prone to failure \cite{Change2TwinBarriers}. More disconcerting is the difficulty of connecting manually configured DTs with external agents, including other DTs \cite{wangHowVastDigital2025}. As the scope of studied systems increases to include more domains, DTs must aggregate into systems of systems, sharing information and models with other DTs \cite{vanbossuytFutureDigitalTwin2025}. The challenge of forming interoperable DTs has been described as one the most significant facing modelers \cite{Budiardjo2021, wangKnowledgegraphbasedMultidomainModel2023}, prompting calls for a universal representation schema for DTs \cite{boschertNextGenerationDigital2018, zhengEmergenceCognitiveDigital2022, ricciWebDigitalTwins2022} that ``is reusable across multiple domains, supports multiple diverse activities, and serves the needs of multiple users'' \cite{NASEMDT}.

\IEEEpubidadjcol

Though many frameworks have been proposed \cite{ISO23247, oshoFourRsFramework2022, drobnjakovicOntologizingDigitalTwin2023, minervaDigitalTwinsProperties2021, itu-tDigitalTwinNetwork2022, longoDistributedSimulationDigital2022, zhengEmergenceCognitiveDigital2022, wangKnowledgegraphbasedMultidomainModel2023, akroydUniversalDigitalTwin2021, waszakLetAssetDecide2022, kapteynProbabilisticGraphicalModel2021, ricciWebDigitalTwins2022}, the authors observe that each is inhibited by the dualism of expressiveness at the cost of interpretability. Knowledge graphs \cite{wangKnowledgegraphbasedMultidomainModel2023, akroydUniversalDigitalTwin2021, waszakLetAssetDecide2022, ricciWebDigitalTwins2022} represent the former end of the spectrum. With their connections defined arbitrarily, knowledge graphs can represent any set of objects and their relationships \cite{chatterjeeComparisonGraphTheoreticApproaches2023}. However, to interpret the meaning of a relationship requires an extensive ontology, restricting their ability to externally integrate \cite{hoganKnowledgeGraphs2022, drobnjakovicOntologizingDigitalTwin2023}. Alternatively, more explicit frameworks may fully convey meaning, but only apply to systems within a singular domain \cite{ISO23247, drobnjakovicOntologizingDigitalTwin2023}, or with limited types of relationships such as Bayesian probability models \cite{kapteynProbabilisticGraphicalModel2021, kapteynDatadrivenPhysicsbasedDigital2022}.

The purpose of this paper is to propose a new framework for DTs that solves the paradox of generalized interpretation by deconstructing system behaviors into the composition of set-based functions. The resulting schema is mathematically rigorous, enables deterministic simulation of systems from and across any domain, and provides mechanisms for ready integration of models with external agents. This framework is demonstrated through building a DT for a microgrid, following the validation methodology of Pedersen et al. \cite{pedersenValidatingDesignMethods2000}. Although the application and formulations of this framework are new, its methods will be familiar to mathematicians and computer scientists versed in the principles of the lambda calculi \cite{barendregtLambdaCalculusIts1981} and functional programming \cite{maclennanFunctionalProgrammingPractice1990}. It is the hope of the authors that applying these principles to DTs will empower scientists and engineers in every field to solve the complex systems framing society's most critical challenges.

\section{Nature of Digital Twins} \label{sec:dtoverview}
The concepts of systems modeling are found in many fields, often resulting in competing interpretations of common words. The purpose of this section is to precisely define how digital twins are interpreted by the authors. These interpretations are specific to this paper and intended to be supplementary in nature; they are not meant to serve as a formal standard or supersede existing definitions provided by works better suited for this purpose \cite{internationalorganizationforstandaridizationDigitalTwinConcepts2023, DigTwinConsortiumDefinition, aiaadigitalengineeringintegrationcommitteeDigitalTwinDefinition2020, grievesVirtuallyIntelligentProduct2019, waggDigitalTwinsStateoftheArt2020, vanbossuytFutureDigitalTwin2025, NASEMDT, shaftoDraftModelingSimulation2010}. To maximize the applicability of the proposed framework, the authors have worked to consider DTs as generally as possible, while maintaining a sufficiently scoped definition of a DT focused on the essential functionalities a DT must provide. This task is made difficult from the verbose range of services ascribed to DTs, such as decision-making \cite{aiaadigitalengineeringintegrationcommitteeDigitalTwinDefinition2020}, receiving \cite{damicoIndustrialInsightsDigital2023} and recording \cite{dattaEmergenceDigitalTwins2017} system data, validating \cite{shaoCredibilityConsiderationDigital2023} and updating \cite{waggDigitalTwinsStateoftheArt2020, hakiriComprehensiveSurveyDigital2024} system models, generating \cite{villalongaDecisionmakingFrameworkDynamic2021} and predicting \cite{vanbossuytFutureDigitalTwin2025} system states, controlling physical systems \cite{michaelIntegrationChallengesDigital2022}, and communicating signals \cite{limStateoftheartSurveyDigital2020, humanDesignFrameworkSystem2023}. It is consequently not always clear which roles should be filled by a DT and which ones are better suited for other, more agentic systems. The remainder of this section focuses on motivating a more strict definition of a DT that allows for more careful consideration of their implementation.

The traditional definition of a DT is commonly given as a virtual representation of a physical system \cite{DigTwinConsortiumDefinition, Grieves2014, shaftoDraftModelingSimulation2010}, used to inform an agent (whether human or automaton) about the state of some system of interest (SOI) in lieu of direct measurement. Their usefulness stems from the enhanced observability of a digital system, where information about the system is obtained with greater ease from the DT than the SOI \cite{NASEMDT}. Exposing system states allows agents to make decisions within the context of the SOI \cite{villalongaDecisionmakingFrameworkDynamic2021}, converting virtualizations into actions influencing the real world. A state is a property of a system that is distinguishable from something else \cite{floridiInformationVeryShort2010}, such as the status of a lightbulb or the fuel level of a generator, and which can vary (or evolve) over a range of values \cite{willemsBehavioralApproachOpen2007}. In this way an agent can distinguish between the lightbulb being \textit{on} or \textit{off}, or fuel level being \textit{full} versus \textit{empty}. The goal of an agent is to observe a set of state values exhibited from the SOI. If all states can be directly measured, then no DT is necessary. This is, however, not predominantly the case, necessitating the construction of a DT that informs the agent where direct observation of the SOI is untenable. The DT performs this through one of two mechanisms: direct coupling, where the DT updates in accordance with corresponding properties of the SOI; or simulation, where the DT evolves under a prescribed manipulation from some initial conditions until it approximates the unmeasured facts \cite{cellierContinuousSystemModeling1991}. The combination of these allow an agent to observe the SOI's state through indirect measurements of the DT. Both mechanisms require a connection from the SOI to the DT enabling the latter to reflect the states of the former \cite{vanbossuytFutureDigitalTwin2025}. The authors posit that a digital system providing these mechanisms, along with the necessary intersystem connection, constitutes a DT. This constitutes a stronger interpretation of DTs that yields to a more rigorous mathematical implementation.

\begin{figure}[tb]
	\centering
	\includegraphics[width=\linewidth]{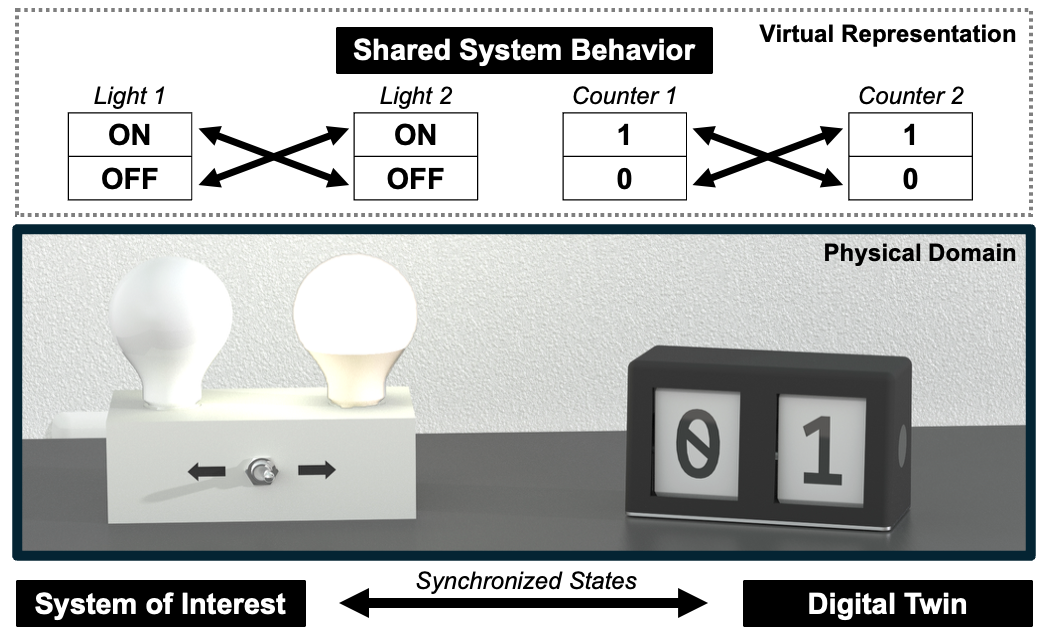}
	\caption{
		A visual description of a DT, with the system of interest and digital twin shown on the left and right respectively in the bottom (physical domain), and a virtual representation of shared behavior of the two systems at the top (virtual domain).}
	\label{fig:lightbulb}
\end{figure}

An example of a DT is introduced in \figurename~\ref{fig:lightbulb} to motivate this refined definition. In the figure is an electrical circuit consisting of two lightbulbs connected by a bimodal switch such that only one lightbulb is illuminated at any instance. The disjointed bulbs form the SOI for some agent who desires to know which light is illuminated. To this end, the agent assigns two values to each lightbulb corresponding to whether the lightbulb is recognized as illuminated (\textit{on}) or not (\textit{off}). If the SOI is fully observable then discovering the current status of each lightbulb is trivial. A more significant case to consider is if the lightbulbs are obscured, preventing direct measurements. A solution involves the construction of the DT shown on the right of \figurename~\ref{fig:lightbulb}, consisting of a box containing a pair of counters each capable of showing boolean integers zero or one. This DT is coupled to the SOI (perhaps via photoresistive sensors) so that the counters correspond to the various states of each lightbulb: one for \textit{on} and zero for \textit{off}. In this context, coupling refers to connecting the signals from one system to another, so that part of the state of one system is shared by the other. The degree of coupling determines whether the prescribed states of the SOI can be identified through observations of the DT.

It is, however, more likely that a SOI cannot be fully coupled to a DT, preventing some facts of the SOI from being exposed to the DT. To exhibit these data, the DT must be configured to have the same behavior as the SOI. Then the SOI and the DT will evolve commensurately, and the facts of the SOI can still be exposed. In the mock example, the behavior of the SOI is shown at the top of \figurename~\ref{fig:lightbulb}, which describes the mutual exclusion of the states for the first and second lightbulbs. This behavior can be mimicked by connecting the boolean counters of the DT so that the left counter only displays the opposite value of the right. After implementing this behavior, the DT needs only to observe a single light to manifest the states of the entire system. Note that DT and SOI do not share any common physical properties: the SOI is electrical; the DT is mechanical, with digital counters. Yet the interpretation of the two systems results in the same virtual specification of a set of states (represented by two boolean variables) and the behavior relating these states.

\section{Representing System Behaviors} \label{sec:behaviors}
This example shows that a DT is a physical, digital system configured in such a way that, when observed, facts of another system can be understood. By necessity, this digital system is programmed to behave similarly to the SOI, so that experimenting on the DT produces the same output as corresponding experiments of the SOI \cite{cellierContinuousSystemSimulation2006}. Any framework proposed for constructing a DT must explicitly reproduce this shared behavior, motivating consideration of how system behavior is defined for simulation.

Simulation is only possible if the agent can discover a relationship between known and unknown facts. These relationships describe a system, and the agent's description of these relationships is a system model \cite{cellierContinuousSystemModeling1991}. Each relationship describes the effect of one variable on another. Willems \cite{willemsBehavioralApproachOpen2007} showed that the specification of all such relationships constitutes a system's behavior, and further that the effect of any behavior can be described as the restriction of the affected variable's possible values. For example, each light bulb in \figurename~\ref{fig:lightbulb} is constrained to only manifest the alternate state of its neighbor. Collecting restrictive relationships together results in an underdetermined model of a virtual system.

Reality is generally much more deterministic \cite{leeDeterminism2021}. Merely reducing a variable's exhibitable values is not indicative of a reified system, which must be temporally consistent, meaning that each variable exhibits only a single value within any frame of consideration \cite{leeDeterminism2021}. To simulate an unknown fact, an agent must discover relationships that reduce the set of possible values for a variable to a single datum. A function is the algebraic mechanism for mapping a value of one set to a value of another, consequently, a simulatable system model is one composed of functions. This can be illustrated by assigning variables to the bimodal lightbulb system in \figurename~\ref{fig:lightbulb}, namely \texttt{light1} and \texttt{light2}, which each can exhibit the states \textit{on} or \textit{off}. The behavior of the first bulb is given by the constraint between \texttt{light2} and \texttt{light1}, described by a function $f$ in Equation~\ref{eq:lightbulb}:

\begin{equation}
	\texttt{light2} \xrightarrow{f} \texttt{light1} := 
	\begin{bmatrix}
		\texttt{on} \\ \texttt{off}
	\end{bmatrix}
	\makebox[-3pt][l]{$\nearrow$}\searrow
	\begin{bmatrix}
		\texttt{on} \\ \texttt{off}
	\end{bmatrix}
	\label{eq:lightbulb}
\end{equation}

As long as the SOI and DT in \figurename~\ref{fig:lightbulb} share states and behavior, equivalent state variables and functions can be applied to the digital display, as in Equation~\ref{eq:counter}, where $g$ is equivalent to $f$.

\begin{equation}
	\texttt{counter2} \xrightarrow{g} \texttt{counter1} :=
	\begin{bmatrix}
		1 \\ 0
	\end{bmatrix}
	\makebox[-3pt][l]{$\nearrow$}\searrow
	\begin{bmatrix}
		1 \\ 0
	\end{bmatrix}
	\label{eq:counter}
\end{equation}

The collection of all known prescriptive functions can be arranged as edges in a graph, with each edge connecting nodes that represent the system variables they relate. In multiple-arity relationships the function maps each combination of the domain variables to a unique value in the codomain. The inclusion of multiple variables (or nodes) in the domain makes the relationship a hyperedge, and the resulting holistic structure is termed a constraint hypergraph (CHG). A technical definition is provided in \cite{morrisUnifiedSystemModeling2025}, which establishes that any explicit system behavior can be represented in a CHG. This is important for a potential DT framework, which must reconcile any models used to stipulate the behavior shared between the DT and SOI into a unified representation. 

\section{Use of Constraint Hypergraphs}
Understanding the structure of a CHG is essential to learning how they fundamentally represent behavior, and consequently to how they can be employed as a universal framework for representing a DT. To this end, this section introduces the high-level information about a CHG, with additional technical information provided in \cite{morrisUnifiedSystemModeling2025}. Although similar to the graphs of constraint programming \cite{rossiConstraintProgramming2008}, CHGs differ in that each edge expresses a complete reduction of a variable's degrees of freedom. The theory of CHGs were originally formalized by \cite{friedmanConstraintTheoryPart1969,friedmanConstraintTheory2017} as a foundation for analyzing systems. Though they have been employed under various names in systems engineering--usually as algebraic links between state variables \cite{peakComposableObjectCOB2005, friedenthalPracticalGuideSysML2015}--the authors assert their usage in representing general system behavior to be far more expansive than previously considered. In addition to the frameworks laid out previously, additional discoveries are presented in this section concerning how CHGs should be configured to represent DTs, specifically: (a) an interpretation of pathfinding; (b) the calculation of cycles in the graph; and (c) how validity frames should be expressed.

\begin{figure*}[htb]
	\centering
	\includegraphics[width=\linewidth]{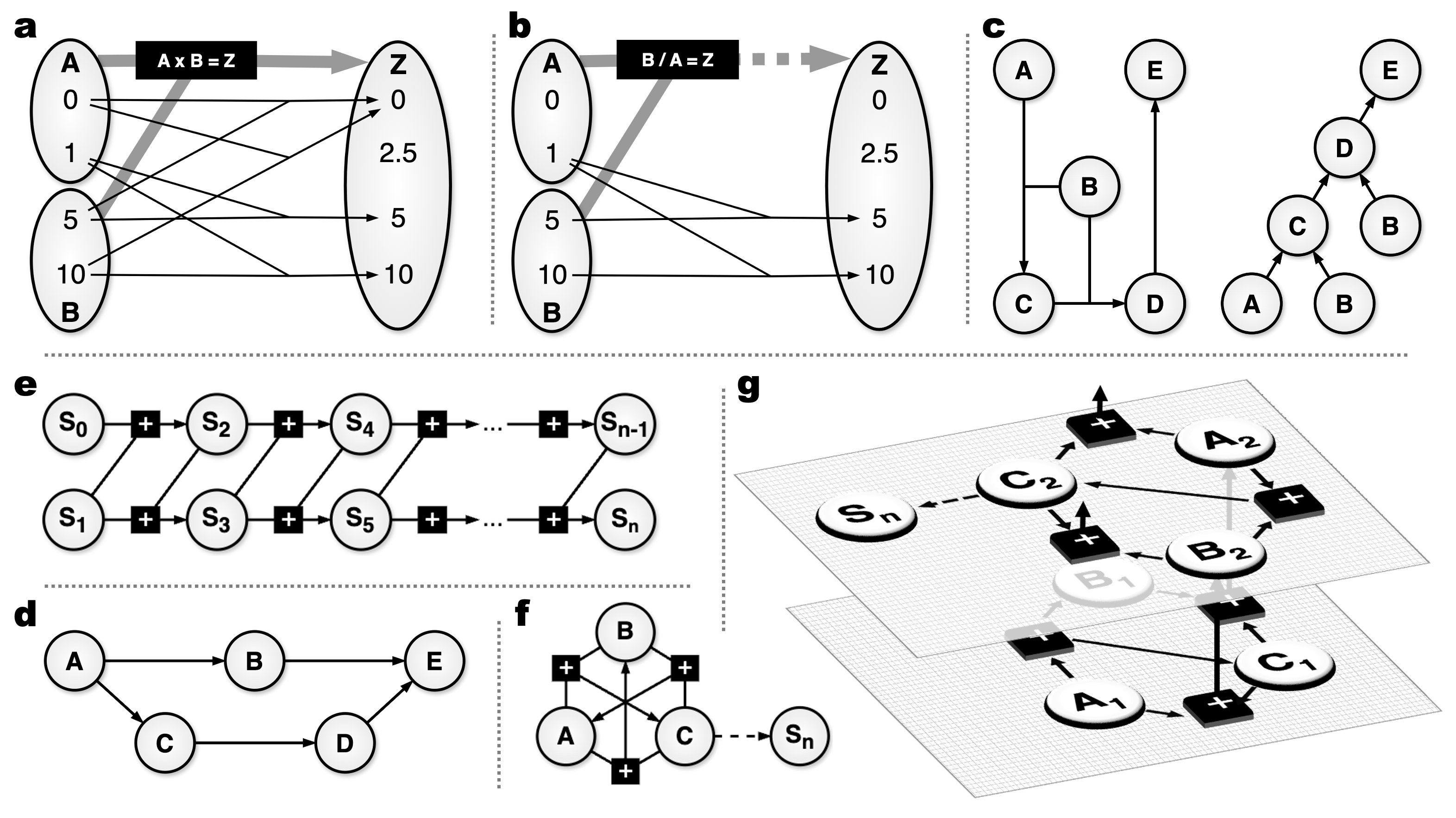}
	\caption{
		Concepts pertaining to CHGs: (\textbf{a}) deconstruction of an edge showing how each combination of set values is mapped to a value in the codomain; (\textbf{b}) similar to \textbf{a} but showing conditional edge mapping from a subset of domain values; (\textbf{c}) a path through a hypergraph (on left) described as a tree (on right); (\textbf{d}) a hypergraph with alternative routes from \textit{A} to \textit{E} demonstrating competing models; (\textbf{e, f, g}) demonstration of cycles using a CHG for computing the $n$-th value of the Fibonacci sequence $S_n$, with an acyclical, distributed CHG in \textbf{e}, and a version using a cycle (and a conditional edge, shown dotted) in \textbf{f}. \textbf{g} shows two iterations of the cyclical CHG in \textbf{f}, with each iteration separated on different planes of connectivity.
		}
	\label{fig:chg_overview}
\end{figure*}

\subsection{Pathfinding}
The structure of a CHG consists of a set of nodes and a set of hyperedges. Each node is a variable and each hyperedge connects a group of nodes (called the source set) to another node (called the target). If the values for each node in the source set are known, the value of the target can be calculated by executing the function encoded in the edge. The target node can then be joined into the source set of another edge, and subsequently used to solve for the value of another variable, in a manner explicitly similar to the lambda calculus foundational to functional languages such as LISP and Haskell \cite{hudakConceptionEvolutionApplication1989}. A path is a chain of edges connecting nodes. The structure of a path through a hypergraph is given by a tree, rooted in a set of source nodes and terminating at a single target \cite{bergeGraphsHypergraphs1973}, as shown in \figurename~\ref{fig:chg_overview}c. Traversing a path is equivalent to simulating the target node and is performed through composing all the functions in the path, as in $(A, B) \xrightarrow{f_1} C; (B, C) \xrightarrow{f_2} D \xrightarrow{f_3} E$ giving $f_3(f_2(B, f_1(A, B))) = E$. The result is a type of universal simulation: the set of variables that can be constrained (and consequently observed) from a set of known values is given by all the nodes reachable by paths rooted in the representative source nodes.

It is possible that multiple paths to a target node exist, as in \figurename~\ref{fig:chg_overview}d. Each path can be interpreted as a competing model capable of solving for the target node. In order to select a preferred model, each path must be compared using some discrimination factor, such as weights given for each edge. Weights can represent criteria such as uncertainty or computational expense, and when summed determine a preferred path \cite{ausielloDirectedHypergraphsIntroduction2017}. 

Each constraint represents an assumption made concerning the relationship between various phenomena, it can consequently be assumed that each constraint is inaccurate to some degree. Establishing edges is done by an independent modeler, and may be performed most conveniently in a domain-specific modeling framework, such as a diagrammatic modeling language. Such frameworks restrict the expressible constraint functions to better capture the domain-specific behavior; geometric kernels, for instance, permit only geometric and dimensional constraints to define a geometric model. The generality of a CHG allows any identified constraint to be recorded in the hypergraph, organizing formerly disparate models into a single, unified structure. Edges encode either a mapping scheme ($a_1 \rightarrow b_1$, $a_2 \rightarrow b_2$, etc., where $a_1, a_2, \ldots$ are the values of a variable $a$) or some calculable rule (e.g. $b = a^2$). A common case is providing a function created by some statistical process such as a neural net or regression algorithm. Functions can even be executable processes, such as calling the Application Programmer Interface (API) of an external client to perform the mapping \cite{morrisDeclarativeSimulationJDSMC2025}. This allows a CHG to connect information across software platforms. In all cases, edges must map each combination of values from a set of variables to a unique datum of another variable. 

\subsection{Conditional Viability}
It is not always true that a behavioral constraint relates all possible states of a system phenomenon. For instance, a model of laminar fluid flow is only valid if the Reynolds number is below a given threshold. Incorporating this model into an edge can be interpreted as the target node being constrained over a subset of the source nodes' values. Instead of representing the subset of the variables over which the constraint is valid as a unique node, a conditional function is supplied for each edge that determines whether the constraint is viable for the known source values (\figurename~\ref{fig:chg_overview}b). 

Conditional viability exacerbates the cost of pathfinding, as paths are no longer independent of inputs to the system. Valid sequences for simulating a given node must be discovered for each new set of inputs, unless it can be shown that the all values solved during the process of traversing the path are within the validity frame of the next edge. Generally, this makes presolving a CHG nonviable, resulting in increased runtime costs during simulation.

\subsection{Cycles}
A cycle is a path in a hypergraph whose target node is also a member of its source set \cite{brettoHypergraphTheoryIntroduction2013}. Cycles should not exist in causal models, as they indicate that a system phenomenon is constrained by its own state. However, there are many situations where system behavior is expressed through a series of repeating patterns, such as a system evolving in time. Each evolution constitutes a unique variable and is consequently represented by a distinct node with distinct edges.

If it can be shown that each instance of a variable is constrained by the same edges, including a dependence upon the previous state of the variable as in \figurename~\ref{fig:chg_overview}e, then the repeated iterations can be summarized into a cycle as in \figurename~\ref{fig:chg_overview}f. Each cycle exists within its own frame of reference, so that all contained nodes are connected only within a single iteration of the cycle. When the cycle returns to the start, the frame of the system is concurrently increased, a pattern visualized in \figurename~\ref{fig:chg_overview}g by a spiral. In order to exit a cycle, a solver must identify an exiting edge that is only viable for the last frame of the cycle.

\subsection{Simulation}
Simulations are defined by a sequence of tasks that, when executed by a digital computer, forms a digital system transforming inputs to outputs (sometimes described as the behavioral trace of a dynamic system \cite{gomesCoSimulationSurvey2018}). In a procedural modeling paradigm, such as block diagrams, each transformation between a set of inputs and an output is encoded in an explicit sequence that can be repeatedly simulated for a range of input values \cite{paredisComposableModelsSimulationBased2001}. However, a unique sequence must be manually defined for each pairing of input to output variables. For a system with $n$ variables, the maximum number of edges that can relate these variables is given by $\sum_{i=1}^{n-1}(n-i)\binom{n}{i}$ \cite{morrisDeclarativeIntegrationEngineering2025}, with each permutation of these edges resulting in a possible simulation. 

Explicitly defining all simulation transformations becomes infeasible even for systems of modest complexity. This is especially critical for DTs, which, to provide appropriate modeling fidelity, typically involve highly-complex system representations extending across multiple scales and domains \cite{NASEMDT}. However, if all processes are not provided, the resulting DT becomes fragmented, failing to manifest the relationships between properties defined by isolated models.  For instance, a finite-element simulation might calculate when a column will buckle, while failing to expose the duration of the buckling event required to execute a discrete-event simulation. Object-oriented frameworks, such as the Functional Mockup-Interface (FMI) \cite{wiensPotentialFMIDevelopment2021}, resolve this by providing an Application-Programmer Interface (API) that specifies the exact information that should be shared between models \cite{piroumianDigitalTwinsUniversal2021}. However, such rigid interfaces result in brittle systems that, while functioning for the systems they are immediately defined for, often fail when the behavior of the SOI changes \cite{krishnamurthiSynthesizingObjectOrientedFunctional1998}. 

\begin{figure}[!htb]
	\includegraphics[width=1.0\linewidth]{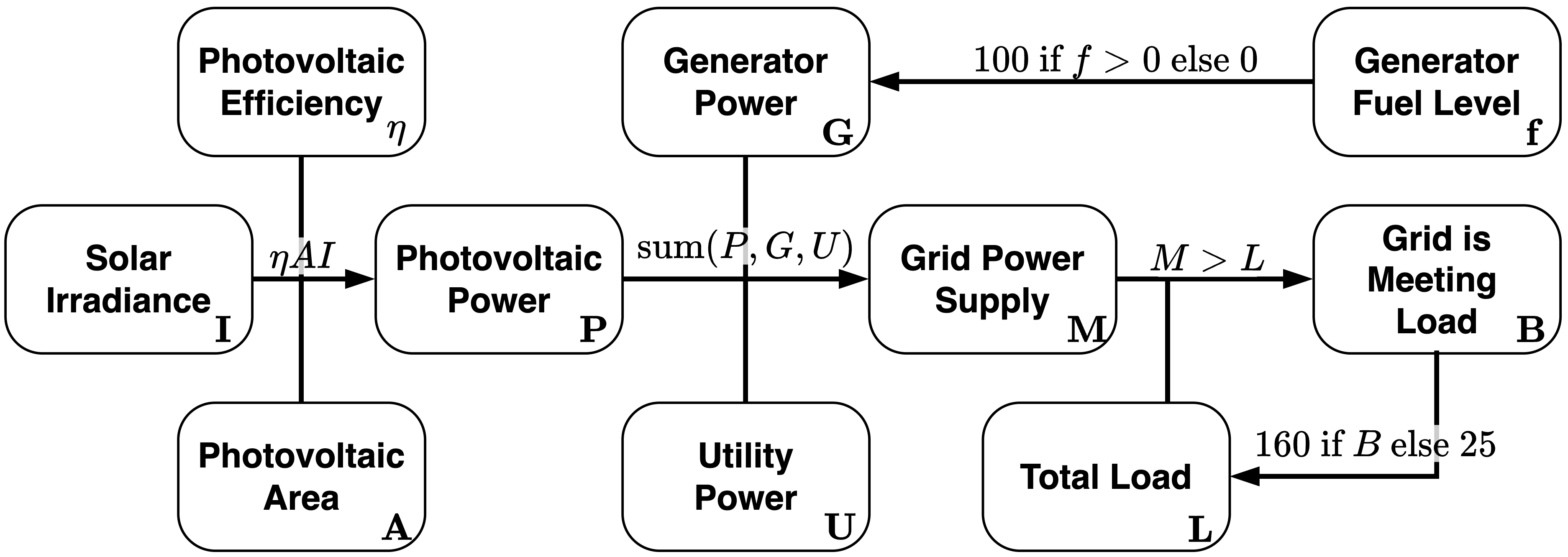}
	\caption{A simple CHG of a microgrid, with nodes for three grid actors (a photovoltaic array $P$, diesel generator $G$, and utility connection $U$) along with associated nodes. The behavioral functions that relate them are given by the arrows (pointing to the target) labeled with the algebraic rule governing the mapping.}
	\label{fig:example-chg}
\end{figure}

In contrast to imperative frameworks, a CHG enables simulation sequences to be autonomously derived from the graphical structure by connecting nodes representing input variables with those representing output variables. The resulting sequence of hyperedges forms a rooted tree \cite{brettoHypergraphTheoryIntroduction2013} composing a function for calculating the desired outputs \cite{morrisDeclarativeSimulationJDSMC2025}. One can think of a CHG as encoding all possible ways a simulation could be composed as paths through the graph. An example of a CHG for a simple microgrid is shown in \figurename~\ref{fig:example-chg}.

\section{Implementation of Digital Twins} \label{sec:practice}
Redefining a DT as a physical system sharing the behavior of another system clarifies what essential actions a framework must provide to implement a DT. In a system whose scope is not changing, the authors posit that there are two primary actions that a DT framework must support: (a) coupling the SOI to the DT, so that the DT has a shared-state with the SOI; and (b) configuring the DT to share the behavior of the SOI. Additionally, if the scope of the twinned system is changing, a framework must additionally (c) provide mechanisms ensuring the twinned representation is valid in the redefined context.

The first action is perhaps the most commonly provided functionality, generally implemented by sensors observing the SOI to the DT, such as an Internet of Things (IoT) network. Such a platform allows measurements of the SOI's state to be reflected in some repository, establishing a fully serviceable DT only if all information of the SOI is collected in the database--a rarity for representations of complex systems. To manifest information that cannot be measured, a DT must also twin the SOI's behavior. This occurs through the provision of models to the DT, which describe how observed inputs may be transformed into unobserved state values. CHGs adeptly perform these two tasks, storing all considered facts as nodes in the systems, and deconstructing models into the edges relating them. A serviceable DT further provides mechanisms for simulating the behavioral models. In addition to fulfilling the first two actions, the primary advantage of using a CHG to represent a DT is this ability to autonomously compose simulation processes. The combination of coupled states and simulatable models allows all desired information of the SOI to be observed via the DT, whether that data is measured or simulated, fulfilling the purpose of a DT. 

While this alone merits consideration of a CHG as a formal framework for DTs, additional advantages derive from their ability to be redefined in different contexts. This last action required for representing a DT--proving a valid representation following arbitrary changes of the SOI's scope--is often characterized as interoperability \cite{Budiardjo2021} or extensibility \cite{duarteProofTheoreticFoundationsDesign1998}, with the challenge of providing it being described the most significant barrier to widespread adoption of DTs \cite{piroumianDigitalTwinsUniversal2021}. Adapting a definition typically employed for software development to a DT, extensibility is the twin's ability to validly manifest a SOI's facts despite changes to the SOI's scope \cite{zengerProgrammingLanguageAbstractions2004}. A model referencing the weight of a vehicle, for instance, might be extensible for changes of the vehicle's simulated location only as long as the location is on the Earth.

Providing extensibility for DTs adds additional complexity due to the expected use of a DT in forming system of systems representations \cite{michaelIntegrationChallengesDigital2022, ricciWebDigitalTwins2022}. Combining DTs into arbitrary aggregate systems is akin to modifying the scope of the original SOI; for example, a DT of a tire and of a car might be combined, expanding the scope of the SOI to the tire + car system. An effective framework ensures the representation of the individual systems remains valid across such scope changes, or else highlights when the representation is potentially invalid. This is especially important given the notion that systems are continuously changing, necessitating DTs that can evolve their virtual representations alongside the physical SOI \cite{waggDigitalTwinsStateoftheArt2020, carlockSystemSystemsSoS2001}.

Decomposing a system into facts and behavioral constraints reveals the two vectors along which system scopes can change: first, adding or removing system data; and second, updating the relationships between that data. The two form the opposing sides of a modeling dilemma referred to as the ``Expression Problem'' \cite{zengerIndependentlyExtensibleSolutions2005, krishnamurthiSynthesizingObjectOrientedFunctional1998}, which describes the difficulty of guaranteeing consistency after arbitrary updates to a model. Famously, a framework can generally be configured to be extensible for changes to either behaviors (relationships) or data, but not both \cite{reynoldsUserDefinedTypesProcedural1978}. Such configured frameworks are respectively described as either functional or procedural \cite{hudakConceptionEvolutionApplication1989}. CHGs are intrinsically functional. This means that, while adding or removing facts in the system can indeed affect a CHG's consistency, modifications of a CHG's edges do not affect the validity of its representation, a characteristic referred to as having no side effects \cite{maclennanFunctionalProgrammingPractice1990}. This is especially useful for DTs, which generally concern an explicitly defined system. Behavioral consistency also allows users to update the models of a DT without needing \textit{a priori} knowledge of the global system, greatly reducing the sensitivity of maintaining a convoluted DT architecture. Examples and further discussion on these points are provided in the Methods section.

\section{Handling Uncertainty} \label{sec:uncertainty}
Uncertainty is inherent and omnipresent in representing reality \cite{boxEmpiricalModelbuildingResponse1987}, making how to reconcile with uncertainty an essential question for any modeling endeavor \cite{klirUncertaintyInformation2006}. As a modeling framework, a CHG neither adds nor removes information to models adhering to its formalisms. However, deconstructing model elements into the nodes and edges of a CHG can provide significant insight into the uncertainty pertaining to a system representation. The tools provided in understanding uncertainty are best understood when framed against the three types of uncertainty described by Isukapalli et al. \cite{isukapalliStochasticResponseSurface1998}: natural variability, model uncertainty, and data uncertainty. This section describes each of these as well as how they can be expressed and understood within the syntax of a CHG.

\begin{figure}[!htb]
	\centering
	\includegraphics[width=\linewidth]{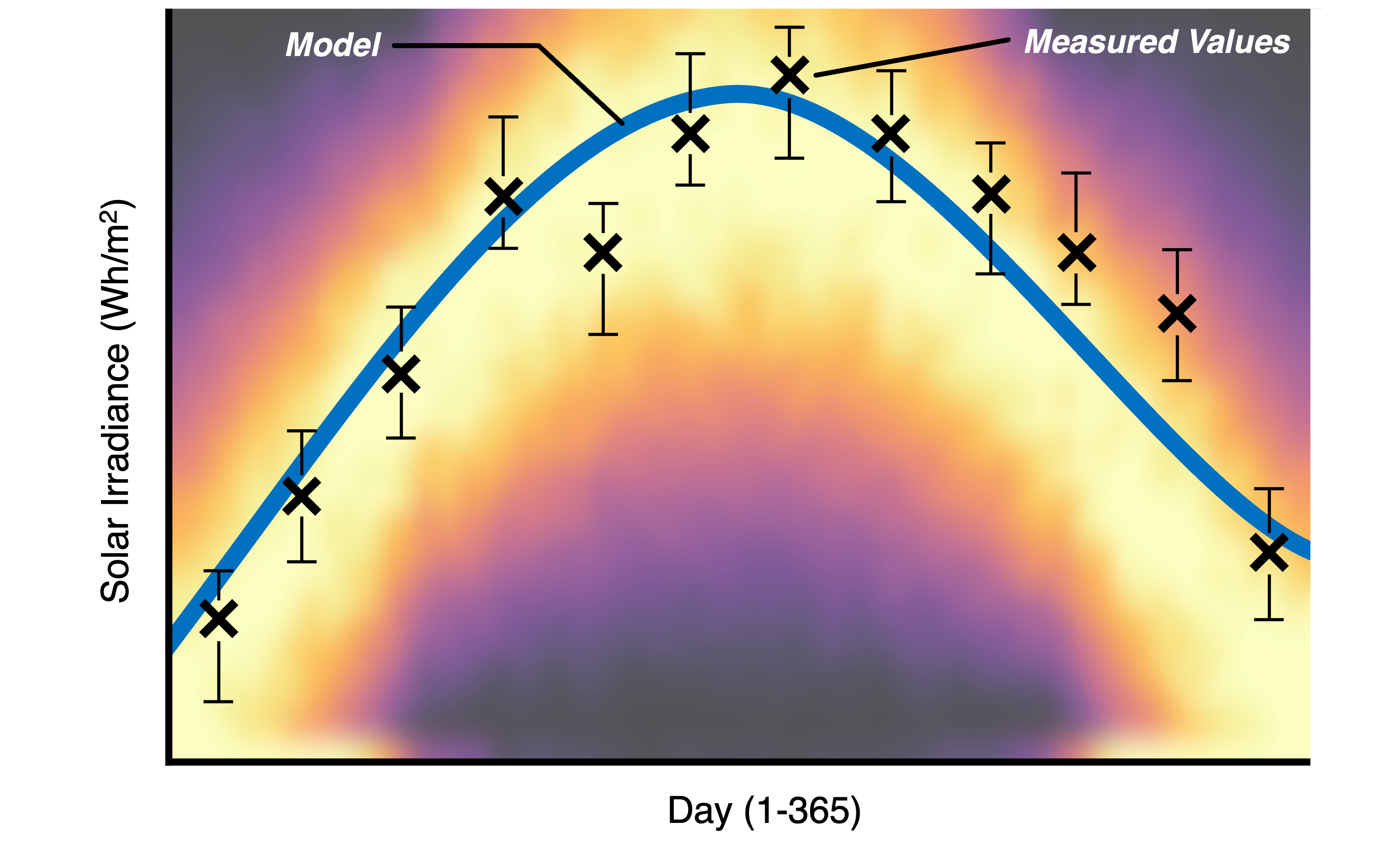}
	\caption{Illustration of modeling uncertainty for a relationship between day of the year and the average solar irradiance. The color map represents the probabilistic value for irradiance, while measured values are given by black X's. A fit model is shown by the blue line. Note that data is given for illustrative purposes only.}
	\label{fig:uncertainty}
\end{figure}

\subsection{Types of Uncertainty}
Natural variability is the uncertainty in a model due to the imperfect approximation of physical phenomenon. As described in Section~\ref{sec:dtoverview}, information is a bounded, finite representation of the physical domain. Such discrete descriptions are inevitably insufficient for capturing the complex interactions of reality. The incongruence between the physical phenomena and its virtual description is intractable, meaning that it cannot be reduced by the efforts of a modeler \cite{choiInductiveDesignExploration2008}. This is described in \figurename~\ref{fig:uncertainty}, which shows an illustrative plot of a model predicting solar irradiance based on the time of year. The relationship between these two phenomena is stochastic rather than deterministic, and is described by the probability density function. This is shown as the color map of the plot where brighter colors are more likely to be measured as real values. No model will perfectly capture this stochastic distribution.

Distinct from natural variation is data uncertainty, which is associated with the fidelity between the inputs provided to the model and the state of the physical system they represent. Each input is a datum that represents the specific measurement of some real world phenomenon. Measurement error, resulting for instance from the resolution of a measuring device or process, forms a key aspect of data uncertainty. An example is given by the measurements of solar irradiance shown by the black Xs in \figurename~\ref{fig:uncertainty}, which cannot be guaranteed to align in the exact position provided. The data uncertainty is described by the error bars describing the tolerance of each measurement.

The final form of uncertainty is model uncertainty, describing unknown information associated with the structure of the model. The formation of every model is based on assumptions. For instance, the interpolation of the measured data points in \figurename~\ref{fig:uncertainty} assumes that the solar irradiance follows a polynomial curve. Assumptions for calculating the energy production of a diesel generator might include the fuel being available and that the combustion cycle performs normatively. Each assumption is an aspect where the model might disagree with the real world, resulting in error. In addition to assumptions in model relationships, the modeler makes assumptions as to how data should be represented. While data uncertainty is the consideration of whether the diesel generator's fuel level is \textit{full} or \textit{empty}, model uncertainty is manifest as the assumption that fuel levels can only be \textit{full} or \textit{empty}. These two forms of uncertainty can be reduced and even eliminated through the efforts of the modeler \cite{sinhaUncertaintyManagementDesign2012}, such as considering states with a higher degree of resolution or making more refined assumptions.

\subsection{Uncertainty in a Constraint Hypergraph}
The structure of a CHG--its nodes and edges--provides a framework for considering uncertainty in a simulation. Nodal elements qualify a system's scope, while their values describe its resolution. Edges represent the assumptions inherent in the model that might need to be accepted to simulate specific information. More specific details can be found by describing how each form of simulation uncertainty is expressed by these two elements.

The fact that natural variability is not reducible relates to its inability to be represented in a model. It is instead the variation inherent in the form of representation selected by the modeler. As such it is not captured within the CHG, though it can be framed through use of the CHG. For instance, although the variability in the relationship shown in \figurename~\ref{fig:uncertainty} between the time of year and solar irradiance cannot be shown in the model, it can be estimated by comparing measured values against repeated simulations run over a sampling space (such as via Monte Carlo methods). These efforts benefit from a CHG, whose structure defines the explicit space for all simulations as well as their construction via the pathfinding measures described in the Methods section.

When an agent prescribes a value for a node they introduce information to the system from outside the model. This represents data uncertainty--the unknown accuracy of an assigned input. Like natural variability, data uncertainty is not captured explicitly in a model, as the information is derived external to the model's scope. However, the flexibility of a CHG allows quantified values of uncertainty, such as the tolerance stack of a measurement process or the confidence interval for a simulation, to be included as a node in the graph representing the specific datum. Treating uncertainty parameters as states in the system reveals the relationships between a model's uncertainty and its simulated outcomes--an important aspect of using models for decision-making and risk estimation.

Contrasted with the first two forms of simulation uncertainty, model uncertainty is intrinsically captured in the structure of the CHG. Nodes encode the set of all distinguishable ways some phenomena can be characterized. As such, each node provides the resolution of the model corresponding to the specific phenomena considered. 

\subsection{Model Assumptions}
Modeling assumptions are captured by the edges in a CHG, such that conducting a simulation implies acceptance of the underlying assumptions for each edge in the simulation path. For instance, \figurename~\ref{fig:microgrid-validation} shows an obvious discrepancy between the simulated and observed outputs of the diesel generators in the validation study. This is due to the assumption present in the DT model that the discharge profile of the diesel generators was uniform, while the controller in the real-world test applied a non-linear reduction at the end of each discharge cycle.

Capturing assumptions affects data uncertainty as well. Every input supplied to a simulation is assumed to true. This is distinct from verifying whether a value is included in an edge's domain--instead it is the assumption that the provided input accurately represents some phenomena. The specific assumption of an input's correctness is captured by a loop: an edge with the same node as its source and target. Loops form the identity element in the category of a CHG, and its existence permits the expression of all modeling assumptions to be expressed solely by the edges in the graph.

\section{Case Study} \label{sec:methods}
\subsection{Overview of Microgrid Demonstration}
\begin{figure}[hb]
	\centering
	\includegraphics[width=\linewidth]{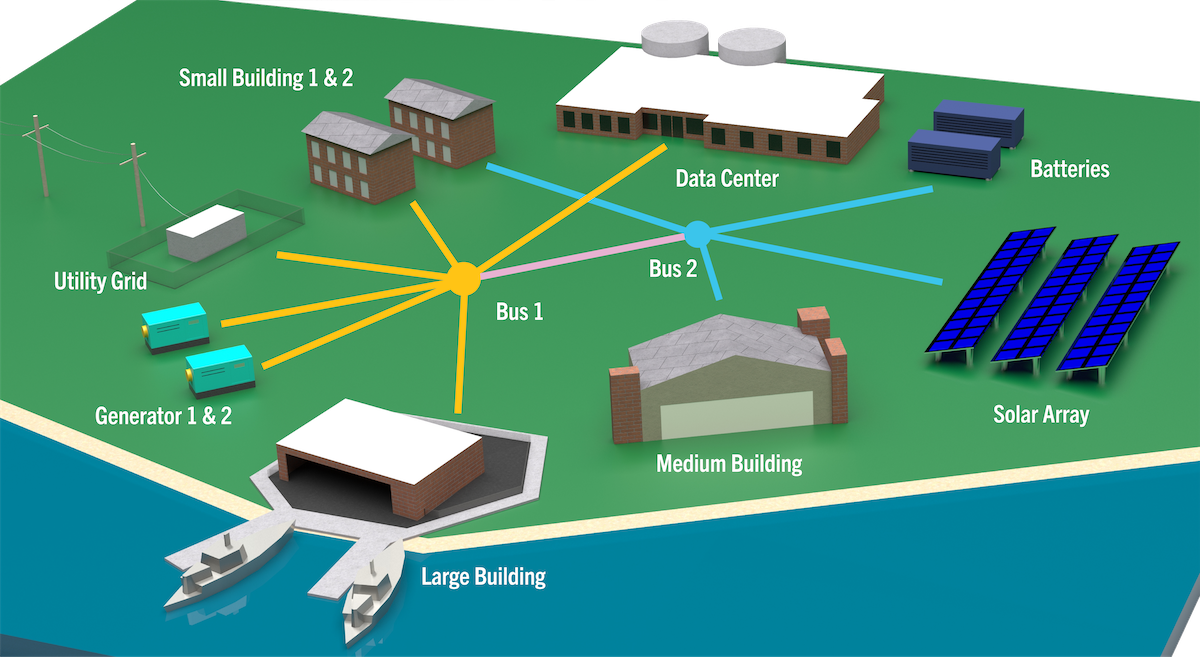}
	\caption{
		Overview of the physical microgrid system, with five suppliers (generators, utility grid, solar cells, and the batteries) and six receivers (buildings, data center, and batteries).}
	\label{fig:microgrid}
\end{figure}

The claim of CHGs being used as a framework for deploying DTs was validated following the methodology of Pedersen et al. \cite{pedersenValidatingDesignMethods2000} by specifically demonstrating a CHG-based DT made for microgrid system. A microgrid is an energy grid that operates semiautonomously from a standard utility network, provide additional flexibility to organizations whose needs go beyond the capabilities of established grids, such as renewable energy producers \cite{lasseterIntegrationDistributedEnergy2002}. Microgrids are composed primarily of energy sources and sinks, here referred to as actors. The system in question, overviewed in \figurename~\ref{fig:microgrid}, consists of an arrangement of grid actors including two diesel generators, a photovoltaic array, a battery energy storage system (BESS), and several energy loads scaled at three possible levels. The system also accounts for connections to the utility grid. Determining the behavior of the microgrid, including the power produced or consumed by its various actors, requires knowledge of loads on the system, the viability of all intermediary connections, and live information on energy fluctuations, such as the availability of solar-generated power. A DT might be used for monitoring purposes as well as testing, with specific importance placed on describing the resilience of the grid to cases of random failure or sabotage \cite{chatterjeeResilientMicrogridDesign2024, andersonjrFoundationsMicrogridResilience2024}. 

\begin{figure}[!htb]
	\centering
	\includegraphics[width=\linewidth]{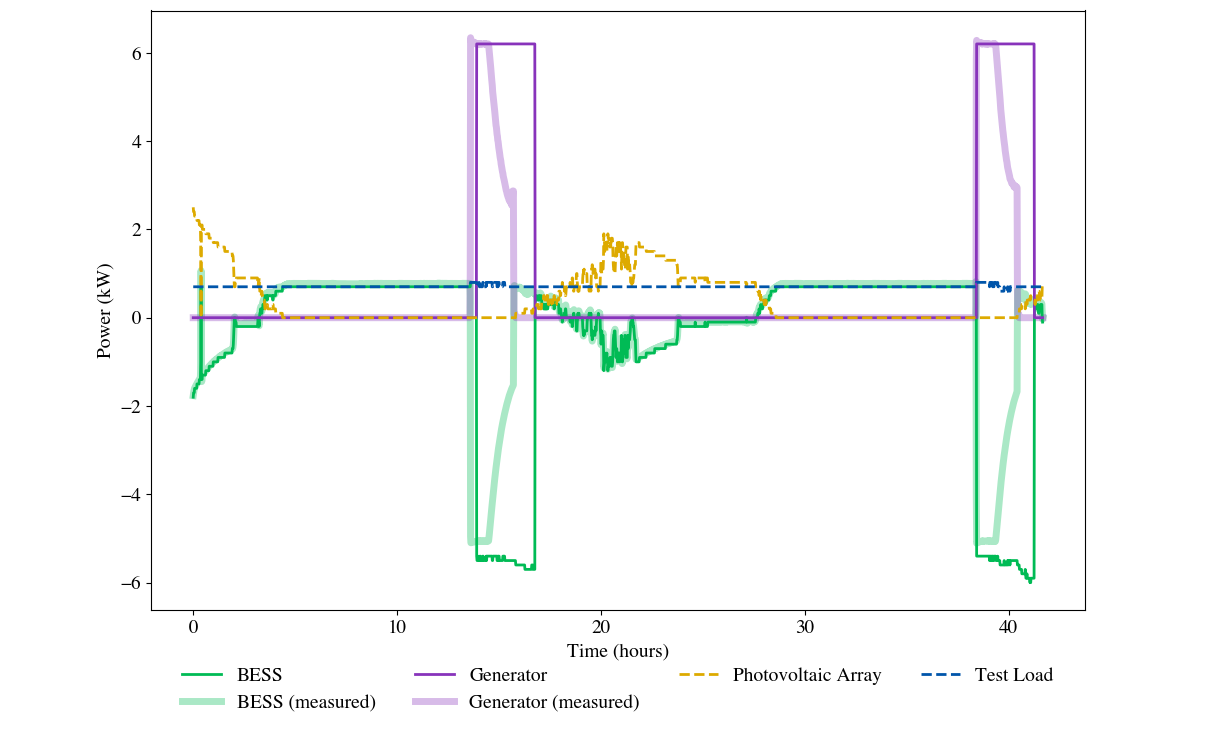}
	\caption{Validation of microgrid DT Model comparing simulated data against a 43-hour test run of a limited microgrid in April 2025 \cite{alvesSpanagelMicrogridExperimental2025}. The dashed lines indicate observed values (directly coupled to the DT), while the solid lines indicate simulated data. Translucent lines indicate the data of the validation run.}
	\label{fig:microgrid-validation}
\end{figure}

The microgrid DT was individually validated by amending it to represent a test-scale microgrid that was operated over a three-day span in Monterey, California in the United States. This bench-top microgrid included a photovoltaic array, a BESS, one diesel generator, and an applied resistive load of approximately 0.7 kW. The DT represented these grid actors by using data inputs from the solar cells and load to simulate the states of the BESS and generator. Comparisons between the test run \cite{alvesSpanagelMicrogridExperimental2025} and simulated values are shown in \figurename~\ref{fig:microgrid-validation}, visually demonstrating its fidelity to the real system.

\subsection{Development}
The broader CHG was derived from standalone, generic models originally developed by researchers associated with the U.S. Navy \cite{oliverosTestModelPower2024, reichMicrogridPlannerOpenSource2024, petersonSystemsArchitectureDesign2019}. These models were developed in largely imperative formats, such as Microsoft Excel workbooks, MATLAB scripts, and Python modules. By deconstructing these into a CHG, the models could be simulated declaratively, rendering an effective DT. The process of conversion from a procedural model to a CHG involved four steps:

\begin{enumerate}
	\item \textbf{Identify facts from the system.} Facts are any piece of information identified in the modeling process, including variables, parameters, or outputted data. These are listed as the nodes of the CHG in-preparation. Facts that were measured from the real system, such as solar irradiance or building usage rates, were set as initial values of their corresponding nodes, establishing the important connection between the virtual model and the physical SOI \cite{waggDigitalTwinsStateoftheArt2020}.
	\item \textbf{Identify relationships between these facts.} Each relationship is described as an edge in the CHG, representing static database queries, discrete-event simulation, linear system solving, stochastic relationships, and file input/output.\footnote{A full, visual overview of the CHG is available on the \href{https://github.com/jmorris335/MicrogridHg/blob/9c9838dceff35012d1266e6b1e40f718224a782d/media/microgrid-chg.png}{online repository} \cite{morrisMicrogridHg2025}.}
	\item \textbf{Organize the model.} The described nodes and edges are parsed and stored into a persistent collection. \label{step:implement}
	\item \textbf{Interrogate the model.} An autonomous agent capable of searching the graph provides observations of the underlying SOI by connecting known inputs with desired outputs along discovered paths in the graph. \label{step:search}
\end{enumerate}

594 variables were identified for Step 1, corresponding to various parameters and states of the microgrid system, as well as settings for the simulation itself.\footnote{Descriptions of all variables are available on the \href{https://github.com/jmorris335/MicrogridHg/blob/9c9838dceff35012d1266e6b1e40f718224a782d/media/table-mg-nodes.png}{online repository} \cite{morrisMicrogridHg2025}.} The collection of all variables forms a state vector, and encodes every dimension along which the system (as described) may evolve. These states blend common interpretations of properties, attributes, and class variables. For instance, the states of a BESS include its current charge level, capacity, but also properties such as its maximum output. If any of these are observable to the modeler--such as from vendor specification sheets--they can be provided as inputs to the CHG, in which case the simulation of that node is the trivial provision of the input node's value (where input equals output). Other properties in the system include system facts conveying simulation settings, data directories, and timing variables. The latter is an interesting aspect of CHGs, where time is considered an observed state of a system, rather than as a universal property (such as in system formalisms given by Wymore \cite{wymoreModelBasedSystemsEngineering1993} or Ziegler \cite{zeiglerTheoryModelingSimulation1976}). The benefit of this is that time can be considered uniquely for composing subsystems. Reconciling two models is contingent upon reconciling time steps and counters, a process captured in reconciling the CHGs.

The edges for each system are based on imperative models written by Peterson \cite{petersonSystemsArchitectureDesign2019} that were converted to a functional structure. The relationships are of a varied nature: some are algebraic, such as the relationship between the number of seconds in a minute and the number of seconds in a year, others are more procedural, such as the functions calculating whether the BESS should be charging based on its charge levels and usage requirements. In every case, functions represent the lowest level of abstraction in the system--a black box within which the solver does not consider system complexity. These black boxes can be incredibly complex if needed; the function relating which grid members should be utilized over a given time step, for example, is calculated via a series of Python methods. By abstracting these steps into a single function, all the sub-variables and relationships in the methods are abstracted away so that they cannot be connected to other nodes in the graph. This is the unfortunate necessity of any system model, which must include a scope beyond which entities are not considered.

\subsection{Use of the Microgrid Digital Twin}
Every path in the hypergraph is a potential simulation pairing a set of known inputs to a unique output. Deploying a CHG following its construction requires an engine capable of storing a CHG, parsing its members, and searching for a path mapping a set of  inputs to an output. Such an engine may accomplish these tasks through any means, with its specific implementation affecting its efficiency of computation. Note that the performance of an engine is only tangentially related to the theoretical proposals here, and is consequently not explored in this study. The engine used by the authors for this article was a custom solver written in the Python programming language titled \textit{ConstraintHg}\cite{morrisConstraintHg2024}, which employs a breadth-first search algorithm. Measured values (system information that is provided as an input) are given by discovering the trivial loop leading from a node to itself. In this case the engine acts as a database-management system: locating the fact's corresponding node and returning the stored value. Non-trivial paths supply simulated information, such as the projected cost of fuel for running the generator. In this case the agent seeds possible paths starting from each input, extending each path (a tree with inputs as its leaves) until one reaches the node corresponding to the desired output. The edges in the path represent the series of functions that, when composed, map the input set of the path's leaves to the path's root. The result of calculating each of these functions is the desired output. 

\begin{figure}[!htb]
	\centering
	\includegraphics[width=\linewidth]{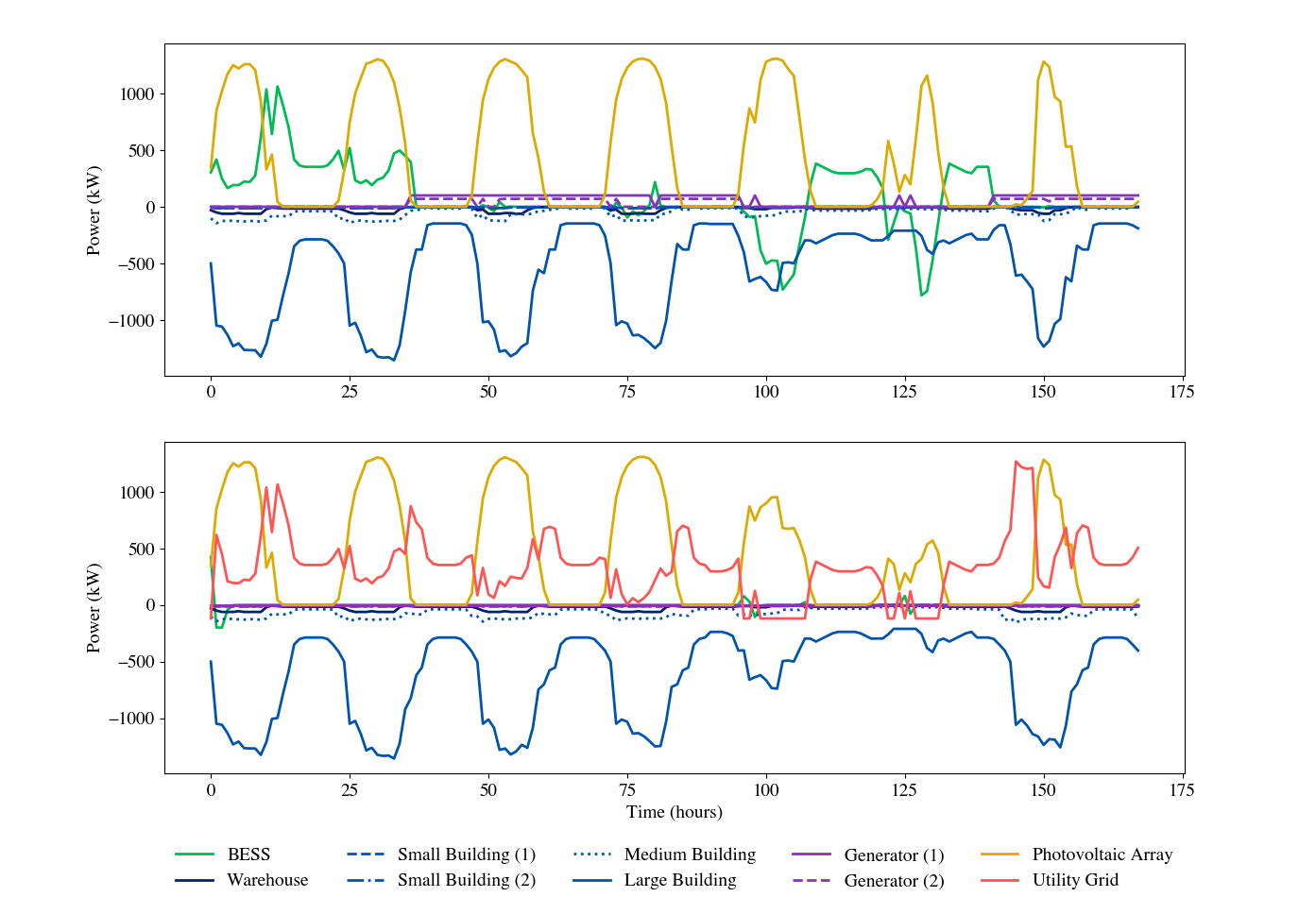}
	\caption{Demonstration of Microgrid CHG showing two simulations of the states of all grid actors over the same week-long span, with the top plot showing states of actors with the utility grid not connected (islanded).}
	\label{fig:microgrid-run}
\end{figure}

After building the CHG and integrating it with the solving engine, the microgrid model was used to demonstrate several possible use cases of a DT:
\begin{enumerate}
	\item \textbf{Data collection:} Relational database querying is perhaps the most similar form of system modeling, where each table represents the functional mapping between sets of values. In this case, the inputs for solar irradiance were taken from data sets compiled by the National Renewable Energy Laboratory (NREL) for Monterey, California, USA over the years 2000 to 2010 \cite{nationalrenewableenergylaboratoryNationalSolarRadiation2012, wilcoxNationalSolarRadiation2012}. All aspects of data querying including file locations, column names, and primary keys were included as nodes in the hypergraph, allowing for full data retrieval and visualization, as shown with the dashed lines in \figurename~\ref{fig:microgrid-validation}.
	
	\item \textbf{Control:} Although the DT was implemented post-data collection as a mock example, feedback control was demonstrated by providing a controller strategy for the generator and BESS systems that prioritized lower-cost energy sources (such as the photovoltaic cells). \figurename~\ref{fig:microgrid-validation} shows the BESS switching from receiving to producing power, as well as the generator simulated as turning on to provide power to the grid when needed. Coupling the digital system to the microgrid actors would enable such real-time feedback to be implemented in the SOI. Methods for implementing such coupling have been widely shown, including programmable logic controllers (PLCs), MQTT messaging services, or built-in microcontrollers.
	
	\item \textbf{System Interrogation:} Interfacing with the DT is the act of observing a state of the SOI. The microgrid DT enables universal interrogation of the microgrid's states by autonomously preparing simulation processes. This is programmatically conducted by calling the \texttt{solve(<state>)} function in the \textit{ConstraintHg} package, passing it the name of the node representing the state to be queried. The takeaway from this is that an agent needs only to know the name of a state and have access to the \texttt{solve} method to be able to make full use of the DT. Serialized data, such as the time series of an actor's power output, can be provided by returning each value solved for along a path. Plots of serialized data are given in \figurename{}~\ref{fig:microgrid-run} and~\ref{fig:microgrid-validation}.
\end{enumerate}

\section{Limitations} \label{sec:limitations}
Many of the limitations of using CHGs to frame DTs are born from the nascent status of the supporting technologies. The custom solver used in this study, \textit{ConstraintHg}, is limited in its solution speed and usability. In particular, the breadth-first search conducted within \textit{ConstraintHg} to compose simulation paths is exhaustive and consequently computationally expensive. However, because these practical limitations are not tied to the theoretical foundations of CHGs, the authors suppose that future developments can address these early issues. Real time monitoring and control with DTs will require a more optimized approach, possibly leveraging search heuristics that allow quicker convergence to the optimal path. 

Another reason limiting the scalability of systems is that pathfinding must be performed for every new input-output pairing. This is because a CHG is only a partial category, such that two functions that share a codomain and domain are not guaranteed to compose. Due to this partiality, a path in the hypergraph is only guaranteed to be solvable for its given set of inputs. For instance, a model within the microgrid for calculating the power received from the utility grid is to check if the utility network is connected to the microgrid. If it is not, then the power received is zero. This model can be explicitly written as an edge $e$ between two nodes $A$ and $B$, the first representing the connection status of the utility grid, and the second one relating the power it is providing. Note that $e$ is only valid if the $A$ is set to be \texttt{False}; there is no mapping provided from $A$ to $B$ by $e$ if $A$ is \texttt{True}. This extends to any path containing $e$. A pathfinding algorithm that encounters $e$ cannot then know whether $e$ can be composed into a valid path until it is supplied with a value for $A$, provided either by simulating an edge connected to $A$ or passed as an input to the search sequence, precluding \textit{a priori} traversal optimization by common methods such as Djikstra's algorithm \cite{ausielloOptimalTraversalDirected1992}. The authors address this in \textit{ConstraintHg} by simulating every encountered edge in the search, so that the domain for every edge is always known to the solver. While this is guaranteed to be convergent, it is undeniably an untenable solution for scaled systems. It is the hope of the authors that this can be addressed through continuing studies into the mathematics of CHGs.

\section{Conclusion} \label{sec:conclusion}
The behavior of a system is a virtual concept, and can be represented as a set of explicit constraint functions acting on the state variables of a system. These functions and variables can be composed into a mathematical structure called a CHG, a declarative model of a system that can be used in constructing analogously-behaving DTs. CHGs do not replace other modeling systems; rather they integrate them into a cohesive, global model. This model can be used to interrogate an SOI, expressing values that were either measured or else calculated within the CHG itself. These cross-cutting measures ensure that DTs built upon CHGs are universally accessible, and generally far more extensible, scalable, and redeployable than traditional DTs. 

This work introduces both theoretical and practical measures for developing DTs for fields as diverse as engineering, organization management, medicine, and ecology. The resulting DTs capture meaning across all domains, providing a mathematically robust framework for representing the complex and integrated systems constituting the world in which we live.

\section*{Data Availability}
The data used to validate the microgrid case study \cite{alvesSpanagelMicrogridExperimental2025} is online at \url{https://doi.org/10.5281/zenodo.15675037}. Data, images, and code for the microgrid digital twin \cite{morrisMicrogridHg2025} are available online at \url{https://doi.org/10.5281/zenodo.15447062}. The software for solving the CHGs \cite{morrisConstraintHg2024} is available online at \url{https://doi.org/10.5281/zenodo.15447063.}

\section*{Acknowledgments}
The authors thank Richard Alves for configuring the Spanagel microgrid \cite{alvesSpanagelMicrogridExperimental2025} used to validate the microgrid DT.

\bibliographystyle{IEEEtran}
\bibliography{preprint}

\begin{thebibliography}{10}
\providecommand{\url}[1]{#1}
\csname url@samestyle\endcsname
\providecommand{\newblock}{\relax}
\providecommand{\bibinfo}[2]{#2}
\providecommand{\BIBentrySTDinterwordspacing}{\spaceskip=0pt\relax}
\providecommand{\BIBentryALTinterwordstretchfactor}{4}
\providecommand{\BIBentryALTinterwordspacing}{\spaceskip=\fontdimen2\font plus
\BIBentryALTinterwordstretchfactor\fontdimen3\font minus \fontdimen4\font\relax}
\providecommand{\BIBforeignlanguage}[2]{{%
\expandafter\ifx\csname l@#1\endcsname\relax
\typeout{** WARNING: IEEEtran.bst: No hyphenation pattern has been}%
\typeout{** loaded for the language `#1'. Using the pattern for}%
\typeout{** the default language instead.}%
\else
\language=\csname l@#1\endcsname
\fi
#2}}
\providecommand{\BIBdecl}{\relax}
\BIBdecl

\bibitem{internationalorganizationforstandardizationSystemsSoftwareEngineering2023}
\BIBentryALTinterwordspacing
{International Organization for Standardization}, ``Systems and {{Software Engineering}} - {{System Life Cycle Processes}},'' May 2023. [Online]. Available: \url{https://www.iso.org/standard/81702.html}
\BIBentrySTDinterwordspacing

\bibitem{sahalPersonalDigitalTwin2022}
R.~Sahal, S.~H. Alsamhi, and K.~N. Brown, ``Personal {{Digital Twin}}: {{A Close Look}} into the {{Present}} and a {{Step Towards}} the {{Future}} of {{Personalised Healthcare Industry}},'' \emph{Sensors}, vol.~22, no.~15, p. 5918, Jan. 2022.

\bibitem{hoffmannDestinationEarthDigital2023}
J.~Hoffmann, P.~Bauer, I.~Sandu, N.~Wedi, T.~Geenen, and D.~Thiemert, ``Destination {{Earth}} -- {{A}} digital twin in support of climate services,'' \emph{Climate Services}, vol.~30, p. 100394, Apr. 2023.

\bibitem{fuDigitalTwinIntegration2022}
Y.~Fu, G.~Zhu, M.~Zhu, and F.~Xuan, ``Digital {{Twin}} for {{Integration}} of {{Design-Manufacturing-Maintenance}}: {{An Overview}},'' \emph{Chinese Journal of Mechanical Engineering}, vol.~35, no.~1, p.~80, Jun. 2022.

\bibitem{koskinenSoftwareMaintenanceCost2003}
\BIBentryALTinterwordspacing
J.~Koskinen, H.~Lahtonen, and T.~Tilus, ``Software {{Maintenance Cost Estimation}} and {{Modernization Support}},'' Information Technology Research Institute, Tech. Rep., Jun. 2003. [Online]. Available: \url{https://static.aminer.org/pdf/PDF/000/364/724/estimating\_the\_costs\_of\_software\_maintenance\_tasks.pdf}
\BIBentrySTDinterwordspacing

\bibitem{zambranoIndustrialDigitalizationIndustry2022}
V.~Zambrano, J.~{Mueller-Roemer}, M.~Sandberg, P.~Talasila, D.~Zanin, P.~G. Larsen, E.~Loeschner, W.~Thronicke, D.~Pietraroia, G.~Landolfi, A.~Fontana, M.~Laspalas, J.~Antony, V.~Poser, T.~Kiss, S.~Bergweiler, S.~Pena~Serna, S.~Izquierdo, I.~Viejo, A.~Juan, F.~Serrano, and A.~Stork, ``Industrial digitalization in the industry 4.0 era: {{Classification}}, reuse and authoring of digital models on {{Digital Twin}} platforms,'' \emph{Array}, vol.~14, p. 100176, 2022.

\bibitem{Change2TwinBarriers}
\BIBentryALTinterwordspacing
P.~Pileggi, A.~Bujari, O.~Barrowclough, J.~Haenisch, and R.~Woitsch, ``Overcoming {{Digital Twin}} barriers for manufacturing {{SMEs}},'' Change2Twin, Position Paper, Apr. 2021. [Online]. Available: \url{https://www.syncontwin.eu/insights/10-overcoming-9-digital-twin-barriers-for-manufacturing-smes/}
\BIBentrySTDinterwordspacing

\bibitem{wangHowVastDigital2025}
X.~Wang, B.~Wu, G.~Zhou, T.~Wang, F.~Meng, L.~Zhou, H.~Cao, and Z.~Tang, ``How a vast digital twin of the {{Yangtze River}} could prevent flooding in {{China}},'' \emph{Nature}, vol. 639, no. 8054, pp. 303--305, Mar. 2025.

\bibitem{vanbossuytFutureDigitalTwin2025}
D.~L. Van~Bossuyt, D.~Allaire, J.~Bickford, T.~A. Bozada, W.~Chen, R.~P. Cutitta, R.~Cuzner, K.~Fletcher, R.~Giachetti, B.~Hale, H.~H. Huang, M.~Keidar, A.~Layton, A.~Ledford, M.~Lesse, J.~Lussier, R.~Malak, B.~Mesmer, G.~Mocko, G.~Oriti, D.~Selva, C.~Turner, M.~Watson, A.~Wooley, and Z.~Zeng, ``The {{Future}} of {{Digital Twin Research}} and {{Development}},'' \emph{Journal of Computing and Information Science in Engineering}, vol.~25, no.~8, p. 080801, Apr. 2025.

\bibitem{Budiardjo2021}
\BIBentryALTinterwordspacing
A.~Budiardjo and D.~Migliori, ``Digital {{Twin System Interoperability Framework}},'' Digital Twin Consortium, Tech. Rep., Dec. 2021. [Online]. Available: \url{https://www.digitaltwinconsortium.org/wp-content/uploads/sites/3/2022/06/Digital-Twin-System-Interoperability-Framework-12072021.pdf}
\BIBentrySTDinterwordspacing

\bibitem{wangKnowledgegraphbasedMultidomainModel2023}
X.~Wang, X.~Hu, Z.~Ren, T.~Tian, and J.~Wan, ``Knowledge-{{Graph-Based Multi-Domain Model Integration Method}} for {{Digital-Twin Workshops}},'' \emph{The International Journal of Advanced Manufacturing Technology}, vol. 128, no. 1-2, pp. 405--421, Sep. 2023.

\bibitem{boschertNextGenerationDigital2018}
\BIBentryALTinterwordspacing
S.~Boschert, C.~Heinrich, and R.~Rosen, ``Next {{Generation Digital Twin}},'' in \emph{Proceedings of {{TMCE}} 2018}.\hskip 1em plus 0.5em minus 0.4em\relax Las Palmas de Gran Canaria, Spain: University of Technology, Delft, May 2018. [Online]. Available: \url{https://www.researchgate.net/publication/325119950\_Next\_Generation\_Digital\_Twin}
\BIBentrySTDinterwordspacing

\bibitem{zhengEmergenceCognitiveDigital2022}
X.~Zheng, J.~Lu, and D.~Kiritsis, ``The emergence of cognitive digital twin: Vision, challenges and opportunities,'' \emph{International Journal of Production Research}, vol.~60, no.~24, pp. 7610--7632, Dec. 2022.

\bibitem{ricciWebDigitalTwins2022}
A.~Ricci, A.~Croatti, S.~Mariani, S.~Montagna, and M.~Picone, ``Web of {{Digital Twins}},'' \emph{ACM Transactions on Internet Technology}, vol.~22, no.~4, pp. 101:1--101:30, Nov. 2022.

\bibitem{NASEMDT}
{National Academies of Sciences, Engineering, and Medicine}, \emph{Foundational {{Research Gaps}} and {{Future Directions}} for {{Digital Twins}}}.\hskip 1em plus 0.5em minus 0.4em\relax Washington, D.C.: The National Academies Press, Mar. 2024.

\bibitem{ISO23247}
\BIBentryALTinterwordspacing
{International Organization for Standardization}, ``Automation systems and integration --- {{Digital}} twin framework for manufacturing,'' Switzerland, Oct. 2021. [Online]. Available: \url{https://www.iso.org/obp/ui/en/\#iso:std:iso:23247:-1:ed-1:v1:en}
\BIBentrySTDinterwordspacing

\bibitem{oshoFourRsFramework2022}
J.~Osho, A.~Hyre, M.~Pantelidakis, A.~Ledford, G.~Harris, J.~Liu, and K.~Mykoniatis, ``Four {{Rs Framework}} for the development of a digital twin: {{The}} implementation of {{Representation}} with a {{FDM}} manufacturing machine,'' \emph{Journal of Manufacturing Systems}, vol.~63, pp. 370--380, Apr. 2022.

\bibitem{drobnjakovicOntologizingDigitalTwin2023}
M.~Drobnjakovic, G.~Shao, A.~Nikolov, B.~Kulvatunyou, S.~P. Frechette, and V.~Srinivasan, ``Towards {{Ontologizing}} a {{Digital Twin Framework}} for {{Manufacturing}},'' in \emph{Advances in {{Production Management Systems}}. {{Production Management Systems}} for {{Responsible Manufacturing}}, {{Service}}, and {{Logistics Futures}}}, ser. {{IFIPAICT}}, vol. 689.\hskip 1em plus 0.5em minus 0.4em\relax Trondheim, NO: Springer, Cham, Sep. 2023, pp. 317--329.

\bibitem{minervaDigitalTwinsProperties2021}
R.~Minerva and N.~Crespi, ``Digital {{Twins}}: {{Properties}}, {{Software Frameworks}}, and {{Application Scenarios}},'' \emph{IT Professional}, vol.~23, no.~1, pp. 51--55, Jan. 2021.

\bibitem{itu-tDigitalTwinNetwork2022}
\BIBentryALTinterwordspacing
{ITU-T}, ``Digital twin network - {{Requirements}} and architecture,'' Feb. 2022. [Online]. Available: \url{https://www.itu.int/rec/T-REC-Y.3090-202202-I}
\BIBentrySTDinterwordspacing

\bibitem{longoDistributedSimulationDigital2022}
F.~Longo, A.~Padovano, L.~Caputi, G.~Gatti, P.~Fragiacomo, V.~D'Augusta, and S.~Talarico, ``Distributed {{Simulation}} for {{Digital Twins}}: An {{Application}} to {{Support}} the {{Autonomous Robotics}} for the {{Extended Ship}},'' in \emph{2022 {{IEEE}}/{{ACM}} 26th {{International Symposium}} on {{Distributed Simulation}} and {{Real Time Applications}} ({{DS-RT}})}, Sep. 2022, pp. 179--186.

\bibitem{akroydUniversalDigitalTwin2021}
J.~Akroyd, S.~Mosbach, A.~Bhave, and M.~Kraft, ``Universal {{Digital Twin}} - {{A Dynamic Knowledge Graph}},'' \emph{Data-Centric Engineering}, vol.~2, p. e14, Jan. 2021.

\bibitem{waszakLetAssetDecide2022}
M.~Waszak, A.~N. Lam, V.~Hoffmann, B.~Elves{\ae}ter, M.~F. Mogos, and D.~Roman, ``Let the {{Asset Decide}}: {{Digital Twins}} with {{Knowledge Graphs}},'' in \emph{2022 {{IEEE}} 19th {{International Conference}} on {{Software Architecture Companion}} ({{ICSA-C}})}, Mar. 2022, pp. 35--39.

\bibitem{kapteynProbabilisticGraphicalModel2021}
M.~G. Kapteyn, J.~V.~R. Pretorius, and K.~E. Willcox, ``A probabilistic graphical model foundation for enabling predictive digital twins at scale,'' \emph{Nature Computational Science}, vol.~1, no.~5, pp. 337--347, May 2021.

\bibitem{chatterjeeComparisonGraphTheoreticApproaches2023}
A.~Chatterjee, C.~Helbig, R.~Malak, and A.~Layton, ``A {{Comparison}} of {{Graph-Theoretic Approaches}} for {{Resilient System}} of {{Systems Design}},'' \emph{Journal of Computing and Information Science in Engineering}, vol.~23, no. 030906, Apr. 2023.

\bibitem{hoganKnowledgeGraphs2022}
A.~Hogan, E.~Blomqvist, M.~Cochez, C.~{d'Amato}, G.~de~Melo, C.~Gutierrez, S.~Kirrane, J.~E.~L. Gayo, R.~Navigli, S.~Neumaier, A.-C. Ngonga~Ngomo, A.~Polleres, S.~M. Rashid, A.~Rula, L.~Schmelzeisen, J.~Sequeda, S.~Staab, and A.~Zimmermann, \emph{Knowledge Graphs}, ser. Synthesis Lectures on Data, Semantics and Knowledge.\hskip 1em plus 0.5em minus 0.4em\relax Cham: Springer, 2022, no.~22.

\bibitem{kapteynDatadrivenPhysicsbasedDigital2022}
{\relax Michael}.~G. Kapteyn, D.~J. Knezevic, D.~B.~P. Huynh, M.~Tran, and {\relax Karen}.~E. Willcox, ``Data-driven physics-based digital twins via a library of component-based reduced-order models,'' \emph{International Journal for Numerical Methods in Engineering}, vol. 123, no.~13, pp. 2986--3003, 2022.

\bibitem{pedersenValidatingDesignMethods2000}
K.~Pedersen, J.~Emblemsv{\aa}g, R.~Bailey, J.~Allen, and F.~Mistree, ``Validating {{Design Methods}} \& {{Research}}: {{The Validation Square}},'' in \emph{{{DETC}} 2000 {{ASME Design Engineering Technical Conferences}}}.\hskip 1em plus 0.5em minus 0.4em\relax Baltimore, MD: ASME, Sep. 2000.

\bibitem{barendregtLambdaCalculusIts1981}
H.~P. Barendregt, \emph{The {{Lambda Calculus}}: {{Its Syntax}} and {{Semantics}}}, ser. Studies in Logic and the Foundations of Mathematics.\hskip 1em plus 0.5em minus 0.4em\relax Amsterdam New York Oxford: North-Holland, 1981, no. 103.

\bibitem{maclennanFunctionalProgrammingPractice1990}
B.~J. MacLennan, \emph{Functional Programming: Practice and Theory}.\hskip 1em plus 0.5em minus 0.4em\relax Reading, Mass: Addison-Wesley, 1990.

\bibitem{internationalorganizationforstandaridizationDigitalTwinConcepts2023}
\BIBentryALTinterwordspacing
{International Organization for Standaridization}, ``Digital {{Twin}}: {{Concepts}} and {{Terminology}},'' Geneva, Switzerland, Nov. 2023. [Online]. Available: \url{https://www.iso.org/standard/81442.html}
\BIBentrySTDinterwordspacing

\bibitem{DigTwinConsortiumDefinition}
\BIBentryALTinterwordspacing
{Digital Twin Consortium}, ``Definition of a {{Digital Twin}},'' Dec. 2020. [Online]. Available: \url{https://www.digitaltwinconsortium.org/initiatives/the-definition-of-a-digital-twin.htm}
\BIBentrySTDinterwordspacing

\bibitem{aiaadigitalengineeringintegrationcommitteeDigitalTwinDefinition2020}
\BIBentryALTinterwordspacing
{AIAA Digital Engineering Integration Committee}, ``Digital {{Twin}}: {{Definition}} and {{Value}},'' {American Institute of Aeronautics and Astronautics and Aerospace Industries Association}, Tech. Rep., Dec. 2020. [Online]. Available: \url{https://www.aiaa.org/docs/default-source/uploadedfiles/issues-and-advocacy/policy-papers/digital-twin-institute-position-paper-(december-2020).pdf}
\BIBentrySTDinterwordspacing

\bibitem{grievesVirtuallyIntelligentProduct2019}
M.~W. Grieves, ``Virtually {{Intelligent Product Systems}}: {{Digital}} and {{Physical Twins}},'' in \emph{Complex {{Systems Engineering}}: {{Theory}} and {{Practice}}}, ser. Progress in {{Astronautics}} and {{Aeronautics}}.\hskip 1em plus 0.5em minus 0.4em\relax {American Institute of Aeronautics and Astronautics, Inc.}, Jan. 2019, vol. 256, pp. 175--200.

\bibitem{waggDigitalTwinsStateoftheArt2020}
D.~J. Wagg, K.~Worden, R.~J. Barthorpe, and P.~Gardner, ``Digital {{Twins}}: {{State-of-the-Art}} and {{Future Directions}} for {{Modeling}} and {{Simulation}} in {{Engineering Dynamics Applications}} [{{Special}} section],'' \emph{ASME J Risk and Uncert in Engrg Sys Part B Mech Engrg}, vol.~6, no.~3, pp. 030\,901:1--030\,901:18, May 2020.

\bibitem{shaftoDraftModelingSimulation2010}
\BIBentryALTinterwordspacing
M.~Shafto, M.~Conroy, R.~Doyle, E.~Glaessgen, C.~Kemp, J.~LeMoigne, and L.~Wang, ``Draft {{Modeling}}, {{Simulation}}, {{Information Technology}} \& {{Processing Roadmap}},'' {National Aeronautics and Space Administration}, Washington DC, Tech. Rep. TA11-27, Nov. 2010. [Online]. Available: \url{https://www.nasa.gov/pdf/501321main\_TA11-MSITP-DRAFT-Nov2010-A1.pdf}
\BIBentrySTDinterwordspacing

\bibitem{damicoIndustrialInsightsDigital2023}
R.~D. D'Amico, S.~Addepalli, and J.~A. Erkoyuncu, ``Industrial {{Insights}} on {{Digital Twins}} in {{Manufacturing}}: {{Application Landscape}}, {{Current Practices}}, and {{Future Needs}},'' \emph{Big Data and Cognitive Computing}, vol.~7, no.~3, p. 126, Sep. 2023.

\bibitem{dattaEmergenceDigitalTwins2017}
S.~P.~A. Datta, ``Emergence of {{Digital Twins}} - {{Is}} this the march of reason?'' \emph{Journal of Innovation Management}, vol.~5, no.~3, pp. 14--33, Nov. 2017.

\bibitem{shaoCredibilityConsiderationDigital2023}
G.~Shao, J.~Hightower, and W.~Schindel, ``Credibility consideration for digital twins in manufacturing,'' \emph{Manufacturing Letters}, vol.~35, pp. 24--28, Jan. 2023.

\bibitem{hakiriComprehensiveSurveyDigital2024}
A.~Hakiri, A.~Gokhale, S.~B. Yahia, and N.~Mellouli, ``A comprehensive survey on digital twin for future networks and emerging {{Internet}} of {{Things}} industry,'' \emph{Computer Networks}, vol. 244, p. 110350, May 2024.

\bibitem{villalongaDecisionmakingFrameworkDynamic2021}
A.~Villalonga, E.~Negri, G.~Biscardo, F.~Castano, R.~E. Haber, L.~Fumagalli, and M.~Macchi, ``A decision-making framework for dynamic scheduling of cyber-physical production systems based on digital twins,'' \emph{Annual Reviews in Control}, vol.~51, pp. 357--373, Jan. 2021.

\bibitem{michaelIntegrationChallengesDigital2022}
J.~Michael, J.~Pfeiffer, B.~Rumpe, and A.~Wortmann, ``Integration challenges for digital twin systems-of-systems,'' in \emph{Proceedings of the 10th {{IEEE}}/{{ACM International Workshop}} on {{Software Engineering}} for {{Systems-of-Systems}} and {{Software Ecosystems}}}, ser. {{SESoS}} '22.\hskip 1em plus 0.5em minus 0.4em\relax New York, NY, USA: Association for Computing Machinery, Nov. 2022, pp. 9--12.

\bibitem{limStateoftheartSurveyDigital2020}
K.~Y.~H. Lim, P.~Zheng, and C.-H. Chen, ``A state-of-the-art survey of {{Digital Twin}}: Techniques, engineering product lifecycle management and business innovation perspectives,'' \emph{Journal of Intelligent Manufacturing}, vol.~31, no.~6, pp. 1313--1337, Aug. 2020.

\bibitem{humanDesignFrameworkSystem2023}
C.~Human, A.~H. Basson, and K.~Kruger, ``A design framework for a system of digital twins and services,'' \emph{Computers in Industry}, vol. 144, p. 103796, Jan. 2023.

\bibitem{Grieves2014}
M.~Grieves, ``Digital {{Twin}}: {{Manufacturing Excellence}} through {{Virtual Factory Replication}},'' Dassault Syst{\`e}mes, Tech. Rep., 2014.

\bibitem{floridiInformationVeryShort2010}
L.~Floridi, \emph{Information: {{A Very Short Introduction}}}, ser. Very {{Short Introductions}}.\hskip 1em plus 0.5em minus 0.4em\relax Oxford: Oxford University Press, 2010.

\bibitem{willemsBehavioralApproachOpen2007}
J.~C. Willems, ``The {{Behavioral Approach}} to {{Open}} and {{Interconnected Systems}},'' \emph{IEEE Control Systems Magazine}, vol.~27, no.~6, pp. 46--99, Dec. 2007.

\bibitem{cellierContinuousSystemModeling1991}
F.~E. Cellier, \emph{Continuous {{System Modeling}}}.\hskip 1em plus 0.5em minus 0.4em\relax New York, NY: Springer, 1991.

\bibitem{cellierContinuousSystemSimulation2006}
F.~E. Cellier and E.~Kofman, \emph{Continuous {{System Simulation}}}.\hskip 1em plus 0.5em minus 0.4em\relax New York: Springer, 2006.

\bibitem{leeDeterminism2021}
E.~A. Lee, ``Determinism,'' \emph{ACM Trans. Embed. Comput. Syst.}, vol.~20, no.~5, pp. 38:1--38:34, May 2021.

\bibitem{morrisUnifiedSystemModeling2025}
J.~Morris, G.~Mocko, and J.~Wagner, ``Unified {{System Modeling}} and {{Simulation}} via {{Constraint Hypergraphs}},'' \emph{Journal of Computing and Information Science in Engineering}, vol.~25, no.~6, p. 061005, Apr. 2025.

\bibitem{rossiConstraintProgramming2008}
F.~Rossi, P.~{van Beek}, and T.~Walsh, ``Constraint {{Programming}},'' in \emph{Foundations of {{Artificial Intelligence}}}, ser. Handbook of {{Knowledge Representation}}, F.~{van Harmelen}, V.~Lifschitz, and B.~Porter, Eds.\hskip 1em plus 0.5em minus 0.4em\relax Elsevier, Jan. 2008, vol.~3, pp. 181--211.

\bibitem{friedmanConstraintTheoryPart1969}
G.~J. Friedman and C.~T. Leondes, ``Constraint {{Theory}}, {{Part II}}: {{Model Graphs}} and {{Regular Relations}},'' \emph{IEEE Transactions on Systems Science and Cybernetics}, vol.~5, no.~2, pp. 132--140, Apr. 1969.

\bibitem{friedmanConstraintTheory2017}
G.~J. Friedman and P.~Phan, \emph{Constraint {{Theory}}}, ser. {{IFSR International Series}} on {{Systems Science}} and {{Engineering}}.\hskip 1em plus 0.5em minus 0.4em\relax Cham: Springer International Publishing, 2017, vol.~23.

\bibitem{peakComposableObjectCOB2005}
\BIBentryALTinterwordspacing
R.~Peak, C.~Paredis, D.~Tamburini, and S.~Waterbury, ``The {{Composable Object}} ({{COB}}) {{Knowledge Representation}}: {{Enabling Advanced Collaborative Engineering Environments}} ({{CEEs}}),'' {National Aeronautics and Space Administration}, Technical {{Report}}, Oct. 2005. [Online]. Available: \url{https://www.eislab.gatech.edu/projects/nasa-ngcobs/COB\_Requirements\_v1.0.pdf}
\BIBentrySTDinterwordspacing

\bibitem{friedenthalPracticalGuideSysML2015}
S.~Friedenthal, A.~Moore, and R.~Steiner, \emph{A {{Practical Guide}} to {{SysML}}: {{The Systems Modeling Language}}}, 3rd~ed.\hskip 1em plus 0.5em minus 0.4em\relax Waltham: Elsevier, 2015.

\bibitem{hudakConceptionEvolutionApplication1989}
P.~Hudak, ``Conception, {{Evolution}}, and {{Application}} of {{Functional Programming Languages}},'' \emph{ACM Comput. Surv.}, vol.~21, no.~3, pp. 359--411, Sep. 1989.

\bibitem{bergeGraphsHypergraphs1973}
C.~Berge, \emph{{Graphs and Hypergraphs}}, ser. {North-Holland Mathematical Library}.\hskip 1em plus 0.5em minus 0.4em\relax Amsterdam, New York: North-Holland Pub. Co., 1973, no.~6.

\bibitem{ausielloDirectedHypergraphsIntroduction2017}
G.~Ausiello and L.~Laura, ``Directed hypergraphs: {{Introduction}} and fundamental algorithms---{{A}} survey,'' \emph{Theoretical Computer Science}, vol. 658, no. Part B, pp. 293--306, Jan. 2017.

\bibitem{morrisDeclarativeSimulationJDSMC2025}
J.~Morris, G.~Mocko, and J.~Wagner, ``Effects of {{Functional}} and {{Declarative Modeling Frameworks}} on {{System Simulation}},'' \emph{Under review with J. Dyn. Sys., Meas., Control (ASME)}, Jun. 2025.

\bibitem{brettoHypergraphTheoryIntroduction2013}
A.~Bretto, \emph{Hypergraph {{Theory}}: {{An Introduction}}}, ser. Mathematical {{Engineering}}.\hskip 1em plus 0.5em minus 0.4em\relax Heidelberg: Springer International Publishing, 2013.

\bibitem{gomesCoSimulationSurvey2018}
C.~Gomes, C.~Thule, D.~Broman, P.~G. Larsen, and H.~Vangheluwe, ``Co-{{Simulation}}: {{A Survey}},'' \emph{ACM Computing Surveys}, vol.~51, no.~3, pp. 49:1--49:33, May 2018.

\bibitem{paredisComposableModelsSimulationBased2001}
C.~Paredis, A.~{Diaz-Calderon}, R.~Sinha, and P.~Khosla, ``Composable {{Models}} for {{Simulation-Based Design}},'' \emph{Engineering with Computers}, vol.~17, no.~2, pp. 112--128, Jul. 2001.

\bibitem{morrisDeclarativeIntegrationEngineering2025}
J.~Morris, A.~Indupally, G.~Mocko, J.~Wagner, and S.~Ramnath, ``Declarative, {{Multi-physics Simulation Between Applications}} via {{Constraint Hypergraphs}},'' \emph{Under review with J. Comput. Inf. Sci. Eng.}, Oct. 2025.

\bibitem{wiensPotentialFMIDevelopment2021}
M.~Wiens, T.~Meyer, and P.~Thomas, ``The {{Potential}} of {{FMI}} for the {{Development}} of {{Digital Twins}} for {{Large Modular Multi-Domain Systems}},'' in \emph{Proceedings of the 14th {{International Modelica Conference}}}, Link{\"o}ping, Sweden, Sep. 2021, pp. 235--240.

\bibitem{piroumianDigitalTwinsUniversal2021}
V.~Piroumian, ``Digital {{Twins}}: {{Universal Interoperability}} for the {{Digital Age}},'' \emph{Computer}, vol.~54, no.~1, pp. 61--69, Jan. 2021.

\bibitem{krishnamurthiSynthesizingObjectOrientedFunctional1998}
S.~Krishnamurthi, M.~Felleisen, and D.~P. Friedman, ``Synthesizing {{Object-Oriented}} and {{Functional Design}} to {{Promote Re-Use}},'' in \emph{Proceedings of the 12th {{European Conference}} on {{Object-Oriented Programming}}}, ser. {{ECCOP}} '98, vol. 1445.\hskip 1em plus 0.5em minus 0.4em\relax Berlin, Heidelberg: Springer-Verlag, Jul. 1998, pp. 91--113.

\bibitem{duarteProofTheoreticFoundationsDesign1998}
C.~H. Duarte, ``Proof-{{Theoretic Foundations}} for the {{Design}} of {{Extensible Software Systems}},'' Ph.D. dissertation, University of London, London, Nov. 1998.

\bibitem{zengerProgrammingLanguageAbstractions2004}
M.~Zenger, ``Programming language abstractions for extensible software components,'' Ph.D. dissertation, EPFL, Lausanne, Switzerland, 2004.

\bibitem{carlockSystemSystemsSoS2001}
P.~G. Carlock and R.~E. Fenton, ``System of {{Systems}} ({{SoS}}) enterprise systems engineering for information-intensive organizations,'' \emph{Systems Engineering}, vol.~4, no.~4, pp. 242--261, 2001.

\bibitem{zengerIndependentlyExtensibleSolutions2005}
\BIBentryALTinterwordspacing
M.~Zenger and M.~Odersky, ``Independently {{Extensible Solutions}} to the {{Expression Problem}},'' in \emph{Foundations of {{Object-Oriented Languages}} ({{FOOL}} 2005)}.\hskip 1em plus 0.5em minus 0.4em\relax Long Beach, CA, USA: {\'E}cole Polytechnique F{\'e}d{\'e}rale de Lausanne, Jan. 2005. [Online]. Available: \url{https://homepages.inf.ed.ac.uk/wadler/fool/program/10.html}
\BIBentrySTDinterwordspacing

\bibitem{reynoldsUserDefinedTypesProcedural1978}
J.~C. Reynolds, ``User-{{Defined Types}} and {{Procedural Data Structures}} as {{Complementary Approaches}} to {{Data Abstraction}},'' in \emph{Programming {{Methodology}}: {{A Collection}} of {{Articles}} by {{Members}} of {{IFIP WG2}}.3}, D.~Gries, Ed.\hskip 1em plus 0.5em minus 0.4em\relax New York, NY: Springer, 1978, pp. 309--317.

\bibitem{boxEmpiricalModelbuildingResponse1987}
G.~E.~P. Box and N.~R. Draper, \emph{Empirical Model-Building and Response Surfaces}, ser. Wiley Series in Probability and Mathematical Statistics.\hskip 1em plus 0.5em minus 0.4em\relax New York: Wiley, 1987.

\bibitem{klirUncertaintyInformation2006}
G.~J. Klir, \emph{Uncertainty and {{Information}}}.\hskip 1em plus 0.5em minus 0.4em\relax Hoboken, NJ, USA: John Wiley \& Sons, Inc, 2006.

\bibitem{isukapalliStochasticResponseSurface1998}
S.~S. Isukapalli, A.~Roy, and P.~G. Georgopoulos, ``Stochastic {{Response Surface Methods}} ({{SRSMs}}) for {{Uncertainty Propagation}}: {{Application}} to {{Environmental}} and {{Biological Systems}},'' \emph{Risk Analysis}, vol.~18, no.~3, pp. 351--363, 1998.

\bibitem{choiInductiveDesignExploration2008}
H.-J. Choi, D.~L. Mcdowell, J.~K. Allen, and F.~Mistree, ``An inductive design exploration method for hierarchical systems design under uncertainty,'' \emph{Engineering Optimization}, vol.~40, no.~4, pp. 287--307, Apr. 2008.

\bibitem{sinhaUncertaintyManagementDesign2012}
A.~Sinha, N.~Bera, J.~K. Allen, J.~H. Panchal, and F.~Mistree, ``Uncertainty {{Management}} in the {{Design}} of {{Multiscale Systems}},'' \emph{Journal of Mechanical Design}, vol. 135, no.~1, p. 011008, Dec. 2012.

\bibitem{lasseterIntegrationDistributedEnergy2002}
R.~Lasseter, A.~Akhil, C.~Marnay, J.~Stephens, J.~Dagle, R.~Guttromsom, A.~S. Meliopoulous, R.~Yinger, and J.~Eto, ``Integration of {{Distributed Energy Resources}}: {{The CERTS Microgrid Concept}},'' Consortium for Electric Reliability Technology Solutions, Berkeley, CA, USA, Consultant {{Report}} LBNL--50829, 799644, Apr. 2002.

\bibitem{chatterjeeResilientMicrogridDesign2024}
A.~Chatterjee, A.~Bushagour, and A.~Layton, ``Resilient {{Microgrid Design Using Ecological Network Analysis}},'' in \emph{The {{Proceedings}} of the 2023 {{Conference}} on {{Systems Engineering Research}}}, D.~Verma, A.~M. Madni, S.~Hoffenson, and L.~Xiao, Eds.\hskip 1em plus 0.5em minus 0.4em\relax Cham: Springer Nature Switzerland, 2024, pp. 603--617.

\bibitem{andersonjrFoundationsMicrogridResilience2024}
W.~W. Anderson~Jr and D.~L. Van~Bossuyt, ``Foundations of {{Microgrid Resilience}},'' in \emph{Microgrids}.\hskip 1em plus 0.5em minus 0.4em\relax John Wiley \& Sons, Ltd, 2024, ch.~28, pp. 655--679.

\bibitem{alvesSpanagelMicrogridExperimental2025}
R.~Alves and D.~Van~Bossuyt, ``Spanagel {{Microgrid Experimental Run}},'' Jun. 2025.

\bibitem{oliverosTestModelPower2024}
\BIBentryALTinterwordspacing
O.~G.~A. Oliveros, ``Test {{Model}} for {{Power Distribution}} on {{U}}.{{S}}. {{Naval Installations}},'' Ph.D. dissertation, Naval Postgraduate School, Monterey, CA, Jun. 2024. [Online]. Available: \url{https://hdl.handle.net/10945/73198}
\BIBentrySTDinterwordspacing

\bibitem{reichMicrogridPlannerOpenSource2024}
D.~Reich and L.~Frye, ``Microgrid {{Planner}}: {{An Open-Source Software Platform}},'' \emph{INFORMS Journal on Computing}, Aug. 2024.

\bibitem{petersonSystemsArchitectureDesign2019}
\BIBentryALTinterwordspacing
C.~J. Peterson, ``Systems {{Architecture Design}} and {{Validation Methods}} for {{Microgrid Systems}},'' Master's thesis, Naval Postgraduate School, Monterey, CA, Sep. 2019. [Online]. Available: \url{https://hdl.handle.net/10945/63493}
\BIBentrySTDinterwordspacing

\bibitem{morrisMicrogridHg2025}
J.~Morris, ``{{MicrogridHg}},'' Clemson, Naval Postgraduate School, May 2025.

\bibitem{wymoreModelBasedSystemsEngineering1993}
A.~W. Wymore, \emph{Model-{{Based Systems Engineering}}: {{An Introduction}} to the {{Mathematical Theory}} of {{Discrete Systems}} and to the {{Tricotyledon Theory}} of {{System Design}}}, ser. Systems Engineering Series.\hskip 1em plus 0.5em minus 0.4em\relax Boca Raton, Fla.: CRC Press, 1993.

\bibitem{zeiglerTheoryModelingSimulation1976}
B.~P. Zeigler, A.~Muzy, and E.~Kofman, \emph{Theory of {{Modeling}} and {{Simulation}}}, 3rd~ed.\hskip 1em plus 0.5em minus 0.4em\relax Elsevier, 1976.

\bibitem{morrisConstraintHg2024}
J.~Morris, ``{{ConstraintHg}},'' Nov. 2024.

\bibitem{nationalrenewableenergylaboratoryNationalSolarRadiation2012}
\BIBentryALTinterwordspacing
{National Renewable Energy Laboratory}, ``National {{Solar Radiation Database}} ({{NSRDB}}),'' https://github.com/jmorris335/MicrogridHg/tree/main/src/solar\_data, Aug. 2012. [Online]. Available: \url{https://nsrdb.nrel.gov/data-viewer}
\BIBentrySTDinterwordspacing

\bibitem{wilcoxNationalSolarRadiation2012}
\BIBentryALTinterwordspacing
S.~Wilcox, ``National {{Solar Radiation Database}} 1991--2010 {{Update}}: {{User}}'s {{Manual}},'' National Renewable Energy Laboratory, Golden, CO, USA, Technical {{Report}} NREL/TP-5500-54824, Aug. 2012. [Online]. Available: \url{https://docs.nrel.gov/docs/fy12osti/54824.pdf}
\BIBentrySTDinterwordspacing

\bibitem{ausielloOptimalTraversalDirected1992}
\BIBentryALTinterwordspacing
G.~Ausiello, R.~Giaccio, G.~Italiano, and U.~Nanni, ``Optimal {{Traversal}} of {{Directed Hypergraphs}},'' International Computer Science Institute, Berkeley, CA, {{ICSI Technical Report}} ICSI TR-92-073, Sep. 1992. [Online]. Available: \url{https://www.icsi.berkeley.edu/icsi/publication\_details?n=778}
\BIBentrySTDinterwordspacing

\end{thebibliography}

\newpage

\section*{Biography Section}
\begin{IEEEbiographynophoto}{John Morris}
	received the B.S. degree from Brigham Young University, Provo, UT, USA, in 2021 and the M.S. degree from Clemson University, Clemson, SC, USA, in 2023. He is currently a Ph.D. candidate at Clemson University in mechanical engineering. His research focuses include systems modeling and simulation, knowledge systems, and industrial methods of digital engineering. He works as the Applications Engineer for the PLM Center at Clemson University. 
\end{IEEEbiographynophoto}

\begin{IEEEbiographynophoto}{Douglas L. Van Bossuyt}
     received the H.B.A. degree in international studies in 2007 and the H.B.S. degree in 2007, the M.S. degree in 2009, and the Ph.D degree in 2012 in mechanical engineering from Oregon State University, Corvallis, OR, USA. He is currently an Associate Professor with the Systems Engineering Department at the Naval Postgraduate School, Monterey, CA, USA and the Director of the Microgrid Innovations Research Center. His research focuses on the nexus of failure and risk analysis, functional modeling and conceptual system design, trade-off studies and decision-making, and resilient systems.
\end{IEEEbiographynophoto}

\begin{IEEEbiographynophoto}{Edward Louis}
	received the B.S. degree in 2020 and M.S. degree in 2022 in mechanical engineering from Clemson University, Clemson, SC, USA. He is currently a Ph.D. candidate at Clemson University. His research interests involve systems engineering, multi-objective optimization, and uncertainty quantification.
\end{IEEEbiographynophoto}

\begin{IEEEbiographynophoto}{Gregory Mocko}
	received the B.S. degree in mechanical engineering and material science from the University of Connecticut, Storrs, CT, USA, in 1999, the M.S. degree in mechanical engineering from Oregon State University, Corvallis, OR, USA, in 2001, and the Ph.D. degree from Georgia Institute of Technology, Atlanta, GA, USA, in 2006. He is currently an Associate Professor of mechanical engineering with Clemson University, Clemson, SC, USA. His research interests include human performance, teaming and communication within complex designs, and knowledge-based manufacturing.
\end{IEEEbiographynophoto}

\begin{IEEEbiographynophoto}{John Wagner}
	(Senior Member, IEEE) received the B.S. and M.S. degrees in mechanical engi- neering from the State University of New York at Buffalo, Buffalo, NY, USA, and the Ph.D. degree from Purdue University, West Lafayette, IN, USA. He is currently the Director of the PLM Center at Clemson University. His research interests include non-linear control theory, diagnostic/prognostic methods, digital engineering processes, and mechatronic system design with application to transportation and power generation systems. He is a fellow of the ASME and SAE.
\end{IEEEbiographynophoto}

\vfill

\end{document}